CrossMark

# Model-based clustering based on sparse finite Gaussian mixtures

**Gertraud Malsiner-Walli · Sylvia Frühwirth-Schnatter · Bettina Grün**

**Abstract** In the framework of Bayesian model-based clustering based on a finite mixture of Gaussian distributions, we present a joint approach to estimate the number of mixture components and identify cluster-relevant variables simultaneously as well as to obtain an identified model. Our approach consists in specifying sparse hierarchical priors on the mixture weights and component means. In a deliberately overfitting mixture model the sparse prior on the weights empties superfluous components during MCMC. A straightforward estimator for the true number of components is given by the most frequent number of non-empty components visited during MCMC sampling. Specifying a shrinkage prior, namely the normal gamma prior, on the component means leads to improved parameter estimates as well as identification of cluster-relevant variables. After estimating the mixture model using MCMC methods based on data augmentation and Gibbs sampling, an identified model is obtained by relabeling the MCMC output in the point process representation of the draws. This is performed using $K$-centroids cluster analysis based on the Mahalanobis distance. We evaluate our proposed strategy in a simulation setup with artificial data and by applying it to benchmark data sets.

## 1 Introduction

Finite mixture models provide a well-known probabilistic approach to model-based clustering. They assume that the data are generated by drawing from a finite set of exchangeable mixture components where each mixture component corresponds to one specific data cluster. More specifically, $N$ observations $\mathbf{y} = (\mathbf{y}_1, \ldots, \mathbf{y}_N)$, $\mathbf{y}_i \in \mathbb{R}^r$, are assumed to be drawn from the following mixture distribution:

$$f(\mathbf{y}_i|\boldsymbol{\theta}_1, \ldots, \boldsymbol{\theta}_K, \boldsymbol{\eta}) = \sum_{k=1}^{K} \eta_k f_k(\mathbf{y}_i|\boldsymbol{\theta}_k), \tag{1}$$

where the mixture components are in general assumed to belong to a well-known parametric distribution family with density $f_k(\mathbf{y}_i|\boldsymbol{\theta}_k)$ and $\boldsymbol{\eta} = (\eta_1, \ldots, \eta_K)$ are the component weights, satisfying $\sum_{k=1}^{K} \eta_k = 1$ and $\eta_k \geq 0$; see McLachlan and Peel (2000) and Frühwirth-Schnatter (2006) for a review of finite mixtures.

Since the pioneering papers by Banfield and Raftery (1993), Bensmail et al. (1997) and Dasgupta and Raftery (1998), model-based clustering based on finite mixtures has been applied successfully in many areas of applied research, such as genetics (McLachlan et al. 2002; Yeung et al. 2001), economics time series analysis (Frühwirth-Schnatter and Kaufmann 2008; Juárez and Steel 2010), social sciences (Handcock et al. 2007), and panel and longitudinal data analysis (McNicholas and Murphy 2010; Frühwirth-Schnatter 2011b), just to mention a few.

G. Malsiner-Walli (✉) · B. Grün
Institut für Angewandte Statistik, Johannes Kepler Universität Linz, Linz, Austria
e-mail: gertraud.malsiner_walli@jku.at

B. Grün
e-mail: bettina.gruen@jku.at

S. Frühwirth-Schnatter
Institute for Statistics and Mathematics, WU
Wirtschaftsuniversität Wien, Vienna, Austria
e-mail: sylvia.fruehwirth-schnatter@wu.ac.at

Despite this success and popularity of model-based clustering based on finite mixtures, several issues remain that deserve further investigation and are addressed in the present paper within a Bayesian framework. Above all, in applications typically little knowledge is available on the specific number of data clusters we are looking for, and, as a consequence, the unknown number $K$ of mixture components corresponding to these data clusters needs to be estimated from the data. Tremendous research effort has been devoted to this question, however, no uniquely best method has been identified. Likelihood-based inference typically relies on model choice criteria such as BIC, approximate weight of evidence, or the integrated classification likelihood criterion to select $K$, see e.g. Biernacki et al. (2000) for a comparison of different criteria. Bayesian approaches sometimes pursue a similar strategy, often adding the DIC to the list of model choice criteria; see e.g. Celeux et al. (2006). However, methods that treat $K$ as an unknown parameter to be estimated jointly with the component-specific parameters are preferable from a Bayesian viewpoint.

Within finite mixtures, a fully Bayesian approach toward selecting $K$ is often based on reversible jump Markov chain Monte Carlo (RJMCMC), as introduced by Richardson and Green (1997). RJMCMC creates a Markov chain that moves between finite mixtures with different number of components, based on carefully selected degenerate proposal densities which are difficult to design, in particular in higher dimensional mixtures, see e.g. Dellaportas and Papageorgiou (2006). Alternatively, the choice of $K$ has been based on the marginal likelihood $p(\mathbf{y}|K)$, which has to be combined with a suitable prior $p(K)$ to obtain a valid posterior distribution $p(K|\mathbf{y})$ over the number $K$ of components (Nobile 2004). However, also the computation of the marginal likelihood $p(\mathbf{y}|K)$ turns out to be a challenging numerical problem in a finite mixture model even for moderate $K$ (Frühwirth-Schnatter 2004).

A quite different approach of selecting the number $K$ of components exists outside the framework of finite mixture models and relies on a nonparametric Bayesian approach based on mixture models with countably infinite number of components. To derive a partition of the data, Molitor et al. (2010) and Liverani et al. (2013) define a Dirichlet process prior on the mixture weights and cluster the pairwise association matrix, which is obtained by aggregating over all partitions obtained during Markov chain Monte Carlo (MCMC) sampling, using partitioning around medoids (PAM; Kaufman and Rousseeuw 1990). The optimal number of clusters is determined by maximizing an associated clustering score.

A second issue to be addressed concerns the selection of cluster-relevant variables, as heterogeneity often is present only in a subset of the available variables. Since the inclusion of unnecessary variables might mask the cluster structure, statistical interest lies in identifying these cluster-relevant variables. Several papers have suggested to solve the selection of the number $K$ of components and the identification of cluster-relevant variables simultaneously. One way is to recast the choice both of $K$ and the cluster-relevant variables as a model selection problem. For instance, in the context of maximum likelihood estimation Raftery and Dean (2006), Maugis et al. (2009) and Dean and Raftery (2010) use a greedy search algorithm by comparing the various models through BIC. Penalized clustering approaches using the $L_1$ norm to shrink cluster means together for variable selection are considered in Pan and Shen (2007), with adaptations of the penalty term taking the group structure into account suggested in Wang and Zhu (2008) and Xie et al. (2008). Based on a model using mode association Lee and Li (2012) propose a variable selection algorithm using a forward search for maximizing an aggregated index of pairwise cluster separability.

In the Bayesian framework, Tadesse et al. (2005) propose RJMCMC techniques to move between mixture models with different numbers of components while variable selection is accomplished by stochastic search through the model space. Stingo et al. (2012) extend their approach in combination with Raftery and Dean (2006) to the discriminant analysis framework. Frühwirth-Schnatter (2011a) pursues a slightly different approach by specifying a normal gamma prior on the component means to shrink the cluster mean for homogeneous components, while model selection with respect to $K$ is performed by calculating marginal likelihoods under these shrinkage priors.

Variable selection in the context of infinite mixture models has also been considered. Kim et al. (2006), for instance, combine stochastic search for cluster-relevant variables with a Dirichlet process prior on the mixture weights to estimate the number of components. In a regression setting Chung and Dunson (2009) and Kundu and Dunson (2014) also use stochastic search variable selection methods in combination with nonparametric Bayesian estimation based on a probit stick-breaking process mixture or a Dirichlet process location mixture. Similarly, Yau and Holmes (2011) define a Dirichlet process prior on the weights, however, they identify cluster-relevant variables by using a double exponential distribution as shrinkage prior on the component means. Lian (2010) uses Dirichlet process priors for simultaneous clustering and variable selection employing a base measure inducing shrinkage on the cluster-specific covariate effects.

The main contribution of the present paper is to propose the use of *sparse finite mixture models* as an alternative to infinite mixtures in the context of model-based clustering. While remaining within the framework of finite mixtures, sparse finite mixture models provide a semiparametric Bayesian approach insofar as neither the number of mixture components nor the cluster-relevant variables are assumed to be known in advance. The basic idea of sparse finite mixture modeling is to deliberately specify an *overfitting* finite mixture model with too many components $K$ and to assume heterogeneity for *all* available variables apriori. Sparse solutions with regard to the number of mixture components and with regard to heterogeneity of component locations are induced by specifying suitable shrinkage priors on, respectively, the mixture weights and the component means. This proposal leads to a simple Bayesian framework where a straightforward MCMC sampling procedure is applied to jointly estimate the unknown number of mixture components, to determine cluster-relevant variables, and to perform component-specific inference.

To obtain sparse solutions with regard to the number of mixture components, an appropriate prior on the weight distribution $\eta = (\eta_1, \ldots, \eta_K)$ has to be selected. We stay within the common framework by choosing a Dirichlet prior on $\eta$, however, the hyperparameters of this prior are selected such that superfluous components are emptied automatically during MCMC sampling. The choice of these hyperparameters is governed by the asymptotic results of Rousseau and Mengersen (2011), who show that, asymptotically, an overfitting mixture converges to the true mixture, if these hyperparameters are smaller than $d/2$, where $d$ is the dimension of the component-specific parameter $\theta_k$.

Sparse finite mixtures are related to infinite mixtures based on a Dirichlet process prior, if a symmetric Dirichlet prior is employed for $\eta$ and the hyperparameter $e_0$ is selected such that $e_0 K$ converges to the concentration parameter of the Dirichlet process as $K$ goes to infinity. For finite $K$, sparse finite mixtures provide a two-parameter alternative to the Dirichlet process prior where, for instance, $e_0$ can be held fixed while $K$ increases.

Following Ishwaran et al. (2001) and Nobile (2004), we derive the posterior distribution of the number of non-empty mixture components from the MCMC output. To estimate the number of mixture components, we derive a point estimator from this distribution, typically, the posterior mode which is equal to the most frequent number of non-empty components visited during MCMC sampling. This approach constitutes a simple and automatic strategy to estimate the unknown number of mixture components, without making use of model selection criteria, RJMCMC, or marginal likelihoods.

Although sparse finite mixtures can be based on arbitrary mixture components, investigation will be confined in the present paper to sparse Gaussian mixtures where the mixture components $f_k(\mathbf{y}_i|\boldsymbol{\theta}_k)$ in (1) arise from multivariate Gaussian densities with component-specific parameters $\boldsymbol{\theta}_k = (\boldsymbol{\mu}_k, \boldsymbol{\Sigma}_k)$ consisting of the component mean $\boldsymbol{\mu}_k$ and the variance-covariance matrix $\boldsymbol{\Sigma}_k$, i.e.

$$f_k(\mathbf{y}_i|\boldsymbol{\theta}_k) = f_{\mathcal{N}}(\mathbf{y}_i|\boldsymbol{\mu}_k, \boldsymbol{\Sigma}_k). \tag{2}$$

To identify cluster-relevant variables within the framework of sparse Gaussian mixtures, we include all variables and assess their cluster-relevance by formulating a sparsity prior on the component means $\boldsymbol{\mu}_k$, rather than excluding variables explicitly from the model as it is done by stochastic search. This strategy to identify cluster-relevant variables has been applied previously by Yau and Holmes (2011) who define a Laplace prior as a shrinkage prior on the mixture component means. To achieve more flexibility and to allow stronger shrinkage, we follow in the present paper Frühwirth-Schnatter (2011a) by using instead the normal gamma prior as a shrinkage prior on the mixture component means which is a two-parameter generalization of the Laplace prior.

Specifying a sparse prior on the component means has in addition the effect of allowing component means to be pulled together in dimensions where the data are homogeneous, yielding more precise estimates of the component means in every dimension. Moreover, the dispersion of the estimated component means in different variables can be compared. In this way, a distinction between cluster-relevant variables, which are characterized by a high dispersion of the cluster locations, and homogeneous variables, where cluster locations are identical, is possible by visual inspection. For high-dimensional data, however, this approach might be cumbersome, as pointed out by a reviewer, and automatic tools for identifying cluster-relevant variables using the posterior distributions of the shrinkage parameters might need to be considered.

Finally, in applied research it is often not only of interest to derive a partition of the data, but also to characterize the clusters by providing inference with respect to the cluster-specific parameters $\boldsymbol{\theta}_k$ appearing in (1). The framework of finite mixtures allows for identification of component-specific parameters, as soon as the label switching problem in an overfitting mixture model with empty components is solved. As suggested by Frühwirth-Schnatter (2001), we ensure balanced label switching during MCMC sampling by adding a random permutation step to the MCMC scheme. For relabeling the draws in a post-processing procedure, a range of different methods has been proposed in the literature, see Sperrin et al. (2010) and Jasra et al. (2005) for an overview. However, most of these proposed relabeling methods become computationally prohibitive for multivariate data with increasing dimensionality and a reasonable number of components.

To obtain a unique labeling we follow Frühwirth-Schnatter (2011a), who suggests to cluster the draws in the point process representation after having removed the draws where the number of non-empty components does not correspond to the estimated number of non-empty components and using only component-specific draws from non-empty components. Clustering the component-specific draws in the point process representation reduces the dimensionality of the relabeling problem, making this method feasible also for mul-tivariate data. For clustering the draws we use $K$-centroids cluster analysis (Leisch 2006) based on the Mahalanobis distance, which allows to fit also elliptical clusters and thereby considerably improves the clustering performance. The cluster assignments for the component-specific draws can be used to obtain a unique labeling and an identified model, if in each iteration each component-specific draw is assigned to a different cluster. We suggest to use this proportion of draws with unique assignment as a qualitative indicator of the suitability of the fitted mixture model for clustering.

The article is organized as follows. Section 2 describes the proposed strategy for selecting the true number of mixture components and introduces the normal gamma prior on the mixture component means. Section 3 provides details on MCMC estimation and sheds more light on the relation between shrinkage on the prior component means and weights. In Sect. 4 the strategy for solving the label switching problem for an overfitting mixture model is presented. In Sect. 5 the performance of the proposed strategy is evaluated in a simulation study and application of the proposed method is illustrated on two benchmark data sets. Section 6 summarizes results and limitations of the proposed approach and discusses issues to be considered in future research.

## 2 Model specification

In a Bayesian approach, the model specification given in Eqs. (1) and (2) is completed by specifying priors for all model parameters. As mentioned in the introduction, sparse finite mixtures rely on specifying a prior on the mixture weights $\boldsymbol{\eta}$ which helps in identifying the number of mixture components (see Sect. 2.1). To achieve identification of cluster-relevant variables, a shrinkage prior on the component means $\boldsymbol{\mu}_1, \ldots, \boldsymbol{\mu}_K$ is chosen (see Sect. 2.2), while a standard hierarchical prior is chosen for the component variances $\boldsymbol{\Sigma}_1, \ldots, \boldsymbol{\Sigma}_K$ (see Sect. 2.3).

### 2.1 Identifying the number of mixture components

Following the usual approach, we assume that the prior on the weight distribution is a symmetric Dirichlet prior, i.e. $\boldsymbol{\eta} \sim Dir(e_0, \ldots, e_0)$. However, since sparse finite mixtures are based on choosing deliberately an overfitting mixture where the number of components $K$ exceeds the true number $K^{true}$, the hyperparameter $e_0$ has to be selected carefully to enable shrinkage of the number of non-empty mixture components toward $K^{true}$.

For an overfitting mixture model, it turns out that the hyperparameter $e_0$ considerably influences the way the posterior distribution handles redundant mixture components. As observed by Frühwirth-Schnatter (2006, 2011a) in an exploratory manner, the posterior distribution of an overfit-ting mixture model with $K > K^{true}$ might exhibit quite an irregular shape, since the likelihood mixes two possible strategies of handling superfluous components. For an overfitting mixture model, high likelihood is assigned either to mixture components with weights close to 0 or to mixture components with nearly identical component-specific parameters. In both cases, several mixture model parameters are poorly identified, such as the component-specific parameters of a nearly empty component in the first case, while only the sum of the weights of nearly identical mixture components, but not their individual values, is identified in the second case.

Rousseau and Mengersen (2011) investigate the asymptotic behavior of the posterior distribution of an overfitting mixture model in a rigorous mathematical manner. They show that the shape of the posterior distribution is largely determined by the size of the hyperparameter $e_0$ of the Dirichlet prior on the weights. In more detail, if the hyperparameter $e_0 < d/2$, where $d$ is the dimension of the component-specific parameter $\boldsymbol{\theta}_k$, then the posterior expectation of the weights asymptotically converges to zero for superfluous components. On the other hand, if $e_0 > d/2$, then the posterior density handles overfitting by defining at least two identical components, each with non-negligible weight. In the second case, the posterior density is less stable than in the first case since the selection of the components that split may vary. Therefore, Rousseau and Mengersen (2011) suggest to guide the posterior towards the first more stable case and to "compute the posterior distribution in a mixture model with a rather large number of components and a Dirichlet-type prior on the weights with small parameters (...) and to check for small weights in the posterior distribution." (p. 694). Following these suggestions, our approach consists in purposely specifying an overfitting mixture model with $K > K^{true}$ being a reasonable upper bound for the number of mixture components. Simultaneously, we favor apriori values of $e_0$ small enough to allow emptying of superfluous components.

An important issue is how to select a specific value for $e_0$ in an empirical application. The asymptotic results of Rousseau and Mengersen (2011) suggest that choosing $e_0 < d/2$ has the effect of emptying all superfluous components, regardless of the specific value, as the number of observations goes to infinity. However, in the case of a finite number of observations, we found it necessary to select much smaller values for $e_0$.

We choose either a very small, but fixed Dirichlet parameter $e_0$, in particular in combination with the sparsity prior on the component means $\boldsymbol{\mu}_k$ introduced in Sect. 2.2, as will be discussed further in Sect. 3.2. Alternatively, to learn from the data how much sparsity is needed, we consider $e_0$ to be an unknown parameter with a gamma hyperprior $\mathcal{G}(a, b)$.

To define the expectation of this prior, we follow Ishwaran et al. (2001) who recommend to choose $e_0 = \alpha/K$. In this case, the Dirichlet prior approximates a Dirichlet process prior with concentration parameter $\alpha$ as $K$ becomes large, as already noted in the introduction. Since simulation studies performed in Ishwaran et al. (2001) yield good approximations for $\alpha = 1$ and $K$ reasonable large, we match the expectation $E(e_0) = 1/K$ obtained in this way:

$$e_0 \sim \mathcal{G}(a, a \cdot K). \tag{3}$$

The parameter $a$ has to be selected carefully since it controls the variance $1/(aK^2)$ of $e_0$. For our simulation studies and applications, we set $a = 10$, as we noted in simulation studies (not reported here) that values smaller than 10 allow large values for $e_0$, and, as a consequence, superfluous components were not emptied during MCMC sampling.

For a sparse finite mixture, the posterior distribution will handle redundant components by assigning to them vanishing weights and, as will be discussed in Sect. 3, superfluous components are emptied during MCMC sampling. Regarding the selection of the number of components, we deviate from Rousseau and Mengersen (2011), because the strategy of separating between "true" and "superfluous" components based on posterior size of the weights of the various components might fail in cases where a threshold for separating between "large" or "small" weights is difficult to identify.

Following, instead, Nobile (2004) and Ishwaran et al. (2001) we derive the posterior distribution $Pr(K_0 = h|\mathbf{y})$, $h = 1, \ldots, K$, of the number $K_0$ of non-empty components from the MCMC output. I.e., for each iteration $m$ of the MCMC sampling to be discussed in Sect. 3, we consider the number of non-empty components, i.e. components to which observations have been assigned for this particular sweep of the sampler,

$$K_0^{(m)} = K - \sum_{k=1}^{K} I\{N_k^{(m)} = 0\}, \tag{4}$$

where $N_k^{(m)}$ is the number of observations allocated to component $k$ and $I$ denotes the indicator function, and estimate the posterior $Pr(K_0 = h|\mathbf{y})$ for each value $h = 1, \ldots, K$, by the corresponding relative frequency.

To estimate the number of mixture components, we derive a point estimator from this distribution. We typically use the posterior mode estimator $\hat{K}_0$ which maximizes the (estimated) posterior distribution $Pr(K_0 = h|\mathbf{y})$ and is equal to the most frequent number of non-empty components visited during MCMC sampling. The posterior mode estimator is optimal under a 0/1 loss function which is indifferent to the degree of overfitting $K_0$. This appears particularly sensible in the present context where adding very small, non-empty components hardly changes the marginal likelihood. This makes the posterior distribution $Pr(K_0 = h|\mathbf{y})$ extremely right-skewed and other point estimators such as the posterior mean extremely sensitive to prior choices, see Nobile (2004).

## 2.2 Identifying cluster-relevant variables

The usual prior on the mixture component means $\boldsymbol{\mu}_k = (\mu_{k1}, \ldots, \mu_{kr})'$ is the independence prior,

$$\boldsymbol{\mu}_k \sim \mathcal{N}(\mathbf{b}_0, \mathbf{B}_0), \quad k = 1, \ldots, K, \tag{5}$$

where $\mathcal{N}(\cdot)$ denotes the multivariate normal distribution. It is common to assume that all component means $\boldsymbol{\mu}_k$ are independent a priori, given data-dependent hyperparameters $\mathbf{b}_0$ and $\mathbf{B}_0$; see e.g. Richardson and Green (1997), Stephens (1997) and Frühwirth-Schnatter (2006). Subsequently, we call this prior the standard prior and choose the median to define $\mathbf{b}_0 = median(\mathbf{y})$ and the range $R_j$ of the data in each dimension $j$ to define $\mathbf{B}_0 = \mathbf{R}_0$, where $\mathbf{R}_0 = \text{Diag}(R_1^2, \ldots, R_r^2)$.

Previous investigations in Yau and Holmes (2011) and Frühwirth-Schnatter (2011a) indicate that it is preferable to replace the standard prior for the component means $\boldsymbol{\mu}_k$ by a shrinkage prior, if it is expected that in some dimensions no cluster structure is present because all component means are homogeneous. Shrinkage priors are well-known from variable selection in regression analysis where they are used to achieve sparse estimation of regression coefficients, see Polson and Scott (2010) and Armagan et al. (2011) for a recent review. Shrinkage priors are also very convenient from a computational point of view, because they can be represented as a scale mixture of normals which makes it easy to implement MCMC sampling under these priors.

We apply in the following the normal gamma prior, for which the mixing distribution for the scale is specified by a gamma distribution. The normal gamma prior was introduced by Griffin and Brown (2010) for variable selection in regression models and has been applied previously by Frühwirth-Schnatter (2011a) in the context of finite mixture distributions. As opposed to the standard prior (5) which is based on fixed hyperparameters $\mathbf{b}_0$ and $\mathbf{B}_0$, a hierarchical prior is introduced, which places a normal prior on the prior mean $\mathbf{b}_0$ and a shrinkage prior on the prior variance matrix $\mathbf{B}_0$:

$$\boldsymbol{\mu}_k|\boldsymbol{\Lambda}, \mathbf{b}_0 \sim \mathcal{N}(\mathbf{b}_0, \mathbf{B}_0), \tag{6}$$

where

$$\mathbf{B}_0 = \boldsymbol{\Lambda} \mathbf{R}_0 \boldsymbol{\Lambda},$$
$$\boldsymbol{\Lambda} = \text{Diag}(\sqrt{\lambda_1}, \ldots, \sqrt{\lambda_r}),$$
$$\lambda_j \sim \mathcal{G}(\nu_1, \nu_2), \quad j = 1, \ldots, r,$$
$$\mathbf{b}_0 \sim \mathcal{N}(\mathbf{m}_0, \mathbf{M}_0).$$

In (6), a multivariate version of the normal gamma prior is employed, where it is assumed that in each dimension $j$ all component means $\mu_{1j}, \ldots, \mu_{Kj}$ follow a normal distribution, where the variance depends on different scaling factors $\lambda_j$ drawn from a gamma distribution with parameters $\nu_1$ and $\nu_2$. The marginal prior for $p(\mu_{1j}, \ldots, \mu_{Kj}|\mathbf{b}_0)$ can be expressed in closed form as (see Frühwirth-Schnatter 2011a):

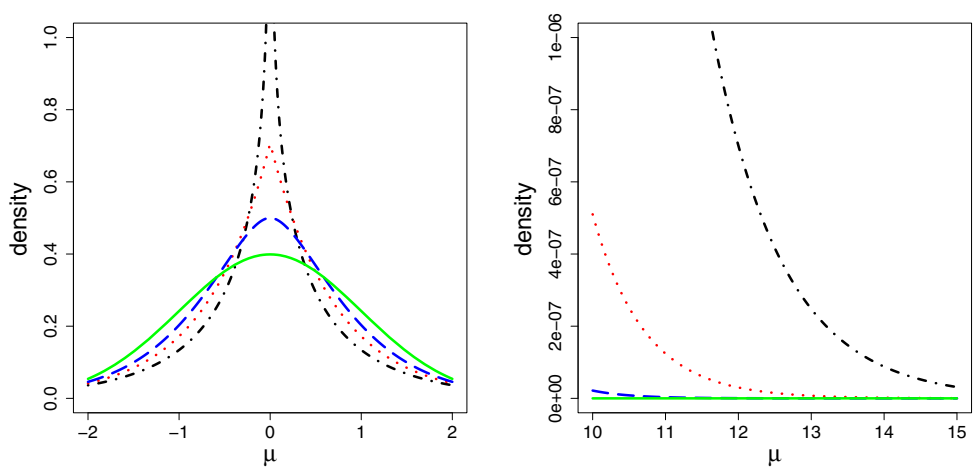

$$\pi(\mu_{1j}, \cdots, \mu_{Kj} | \mathbf{b}_0)$$
$$= \frac{v_2^{v_1}}{(2\pi)^{K/2} \Gamma(v_1)} 2 K_{p_K}(\sqrt{a_j b_j}) \left(\frac{b_j}{a_j}\right)^{p_K/2}, \qquad (7)$$

where

$$a_j = 2v_2,$$
$$p_K = v_1 - K/2,$$
$$b_j = \sum_{k=1}^{K} (\mu_{kj} - b_{0j})^2 / R_j^2,$$

and $K_\alpha(x)$ is the modified Bessel function of the second kind. Furthermore, if the hyperparameters $v_1$ and $v_2$ are equal, then in each dimension $j$ the marginal variance of $\mu_{kj}$ is equal to $R_j^2$ as for the standard prior.

Yau and Holmes (2011) considered a closely related, special case of prior (6) where $v_1 = 1$, which corresponds to the double exponential or Laplace prior, also known as the Bayesian Lasso (Park and Casella 2008). However, in the context of regression analysis, this specific prior has been shown to be suboptimal in the sense that shrinkage to 0 is too weak for insignificant coefficients, while a bias is introduced for large coefficients, see e.g. Polson and Scott (2010) and Armagan et al. (2011). The normal gamma prior introduced by Griffin and Brown (2010) is more flexible in this respect. Since the excess kurtosis is given by $3/v_1$, the normal gamma prior has heavier tails than the Laplace distribution for $v_1 < 1$, reducing the bias for large coefficients. At the same time, it is more peaked than the Laplace distribution which leads to stronger shrinkage to 0 for insignificant coefficients. This can be seen in Fig. 1 where the normal gamma prior is plotted for $v_1 = 0.5, 1, 2$ and compared to the standard normal distribution.

In the context of finite mixtures, the normal gamma prior introduces exactly the flexibility we are seeking to identify cluster-relevant variables. To achieve this goal, the normal gamma prior is employed in (6) with value $v_1 < 1$. This implies that $\lambda_j$ can assume values considerable smaller than 1 in dimension $j$, which leads the prior distribution of $\mu_{kj}$ to concentrate around the mean $b_{0j}$, pulling all the component means $\mu_{kj}$ towards $b_{0j}$. This property becomes important in dimensions where component densities are heavily overlapping or in the case of homogeneous variables, where actually no mixture structure is present and all observations are generated by a single component only. In these cases, allowing the prior variance to pull component means together yields more precise estimates of the actually closely adjacent or even identical component means.

In this way implicit variable selection is performed and variables which are uninformative for the cluster structure are effectively fit by a single component avoiding overfitting heterogeneity and diminishing the masking effect of these variables. The same benefits regarding the fitted model are obtained as if the variables were excluded through a model search procedure.

For cases where the variance of the prior for $\mu_{kj}$, $k = 1, \ldots, K$, is shrunk to a small value, the mean $b_{0j}$ of the prior becomes important. Thus, rather than assuming that $\mathbf{b}_0$ is a fixed parameter as for the standard prior, we treat $\mathbf{b}_0$ as an unknown hyperparameter with its own prior distribution, see (6).

While variable selection is performed only implicitly with shrinkage priors in Bayesian estimation, explicit identification of the relevant clustering variables is possible a posteriori for the hierarchical shrinkage prior based on the normal gamma distribution. In the context of multivariate finite mixtures, $\lambda_j$ can be interpreted as a local shrinkage factor in dimension $j$ which allows *a priori* that component means $(\mu_{1j}, \ldots, \mu_{Kj})$ are pulled together and, at the same time, is flexible enough to be overruled by the data *a posteriori*, if the component means are actually different in dimension $j$. Hence, a visual inspection of the *posterior* distributions of the scaling factors $\lambda_j$, $j = 1, \ldots, r$, e.g. through box plots as in Yau and Holmes (2011), reveals in which dimension $j$ a high dispersion of the component means is present and where, on the contrary, all component means are pulled together.

It remains to discuss the choice of the hyperparameters $\nu_1$, $\nu_2$, $\mathbf{m}_0$ and $\mathbf{M}_0$ in (6). In the following simulation studies and applications, the hyperparameters $\nu_1$ and $\nu_2$ are set to 0.5 to allow considerable shrinkage of the prior variance of the component means. Furthermore, we specify an improper prior on $\mathbf{b}_0$, where $\mathbf{m}_0 = median(\mathbf{y})$ and $\mathbf{M}_0^{-1} = \mathbf{0}$.

### 2.3 Prior on the variance-covariance matrices

Finally, a prior on the variance-covariance matrices $\mathbf{\Sigma}_k$ has to be specified. Several papers, including Raftery and Dean (2006) and McNicholas and Murphy (2008), impose constraints on the variance-covariance matrices to reduce the number of parameters which, however, implies that the correlation structure of the data needs to be modeled explicitly.

In contrast to these papers, we do not focus on modeling sparse variance-covariance matrices, we rather model the matrices without constraints on their geometric shape. Following Stephens (1997) and Frühwirth-Schnatter (2006, p. 193), we assume the conjugate hierarchical prior $\mathbf{\Sigma}_k^{-1} \sim \mathcal{W}(c_0, \mathbf{C}_0)$, $\mathbf{C}_0 \sim \mathcal{W}(g_0, \mathbf{G}_0)$, where $\mathcal{W}(\cdot)$ denotes the Wishart distribution. Regularization of variance-covariance matrices in order to avoid degenerate solutions is achieved through specification of the prior hyperparameter $c_0$ by choosing

$$c_0 = 2.5 + \frac{r-1}{2},$$
$$g_0 = 0.5 + \frac{r-1}{2},$$
$$\mathbf{G}_0 = \frac{100 g_0}{c_0} \, \mathrm{Diag}(1/R_1^2, \ldots, 1/R_r^2),$$

see Frühwirth-Schnatter (2006, p. 192).

## 3 Bayesian estimation

To cluster $N$ observations $\mathbf{y} = (\mathbf{y}_1, \ldots, \mathbf{y}_N)$, it is assumed that the data are drawn from the mixture distribution defined in (1) and (2), and that each observation $\mathbf{y}_i$ is generated by one specific component $k$.

The corresponding mixture likelihood derived from (1) and (2) is combined with the prior distributions introduced, respectively, for the weights $\mathbf{\eta}$ in Sect. 2.1, for the component means $\mathbf{\mu}_k$ in Sect. 2.2, and for $\mathbf{\Sigma}_k$ in Sect. 2.3, assuming independence between these components. The resulting posterior distribution does not have a closed form and MCMC sampling methods have to be employed, see Sect. 3.1.

The proposed strategy of estimating the number of components relies on the correct identification of non-empty components. In Sect. 3.2 we study in more detail that prior dependence between $\mathbf{\mu}_k$ and $\mathbf{\eta}$ might be necessary to achieve this goal. In particular, we argue why stronger shrinkage of very small component weights $\eta_k$ toward 0 might be necessary for the normal gamma prior (6) than for the standard prior (5), by choosing a very small value of $e_0$.

### 3.1 MCMC sampling

Estimation of the sparse finite mixture model is performed through MCMC sampling based on data augmentation and Gibbs sampling (Diebolt and Robert 1994; Frühwirth-Schnatter 2006, chap. 3). To indicate the component from which each observation stems, latent allocation variables $\mathbf{S} = (S_1, \ldots, S_N)$ taking values in $\{1, \ldots, K\}^N$ are introduced such that

$$f(\mathbf{y}_i | \mathbf{\theta}_1, \ldots, \mathbf{\theta}_K, S_i = k) = f_{\mathcal{N}}(\mathbf{y}_i | \mathbf{\mu}_k, \mathbf{\Sigma}_k), \tag{8}$$

and

$$Pr(S_i = k | \mathbf{\eta}) = \eta_k. \tag{9}$$

As suggested by Frühwirth-Schnatter (2001), after each iteration an additional random permutation step is added to the MCMC scheme which randomly permutes the current labeling of the components. Random permutation ensures that the sampler explores all $K!$ modes of the full posterior distribution and avoids that the sampler is trapped around a single posterior mode, see also Geweke (2007). Without the random permutation step, it has to be verified for each functional of the parameters of interest, whether it is invariant to relabeling of the components. Only in this case, it does not matter whether the random permutation step is performed. The detailed sampling scheme is provided in Appendix 1 and most of the sampling steps are standard in finite mixture modeling, with two exceptions.

The first non-standard step is the full conditional distribution $p(\lambda_j | \mu_{1j}, \ldots, \mu_{Kj}, \mathbf{b}_0)$ of the shrinkage factor $\lambda_j$. The combination of a gamma prior for $\lambda_j$ with the product of $K$ normal likelihoods $p(\mu_{kj} | \lambda_j, b_{0j})$, where the variance depends on $\lambda_j$, yields a generalized inverted Gaussian distribution ($\mathcal{GIG}$) as posterior distribution, see Frühwirth-Schnatter (2011a). Hence,

$$p(\lambda_j | \mu_{1j}, \ldots, \mu_{Kj}, \mathbf{b}_0) \sim \mathcal{GIG}(a_j, b_j, p_K),$$

where the parameters $a_j$, $b_j$, and $p_K$ are defined in (7).

Furthermore, if the hyperparameter $e_0$ of the Dirichlet prior is random, a random walk Metropolis-Hastings step is implemented to sample $e_0$ from $p(e_0 | \mathbf{\eta})$, where

$$p(e_0 | \mathbf{\eta}) \propto p(e_0) \frac{\Gamma(K e_0)}{\Gamma(e_0)^K} \left( \prod_{k=1}^{K} \eta_k \right)^{e_0 - 1}, \tag{10}$$

and $p(e_0)$ is equal to the hyperprior introduced in (3).

### 3.2 On the relation between shrinkage in the weights and in the component means

As common for finite mixtures, MCMC sampling alternates between a classification and a parameter simulation step, see Appendix 1. During classification, observations are allocated to component $k$ according to the (non-normalized) conditional probability $\eta_k f_\mathcal{N}(\mathbf{y}_i | \boldsymbol{\mu}_k, \boldsymbol{\Sigma}_k)$, whereas the component-specific parameters $\boldsymbol{\mu}_k$, $\boldsymbol{\Sigma}_k$ and the weight $\eta_k$ are simulated conditional on the current classification $\mathbf{S}$ during parameter simulation. If no observation is allocated to component $k$ during classification, then, subsequently, all component-specific parameters of this empty component are sampled from the prior. In particular, the location $\boldsymbol{\mu}_k$ of an empty component heavily depends on the prior location $\mathbf{b}_0$ and prior covariance matrix $\mathbf{B}_0$.

Under the standard prior (5), where

$$\mathbf{B}_0 = \mathrm{Diag}(R_1^2, \ldots, R_r^2),$$

the location $\boldsymbol{\mu}_k$ of an empty component is likely to be far away from the data center $\mathbf{b}_0$, since in each dimension $j$ with 5 % probability the $\mu_{kj}$ will be further away from $b_{0j}$ than $2 \cdot R_j$. As a consequence, $f_\mathcal{N}(\mathbf{y}_i | \boldsymbol{\mu}_k, \boldsymbol{\Sigma}_k)$ is very small for any observation $\mathbf{y}_i$ in the subsequent classification step and an empty component is likely to remain empty under the standard prior.

In contrast, under the normal gamma prior (6), where $\mathbf{B}_0 = \mathrm{Diag}(R_1^2 \cdot \lambda_1, \ldots, R_r^2 \cdot \lambda_r)$, the scaling factor $\lambda_j$ shrinks the prior variance of $\mu_{kj}$ considerably, in particular in dimensions, where the component means are homogeneous. However, the scaling factor $\lambda_j$ adjusts the prior variance also in cluster-relevant dimensions, since $R_j^2$ is generally much larger than the spread of the non-empty component means which are typically allocated within the data range $R_j$. As a consequence, the location $\mu_{kj}$ of an empty component is either close to the data center $b_{0j}$ (in the case of homogeneous variables) or close to the range spanned by the locations of the non-empty components (in the case of cluster-relevant variables). In both cases, evaluating $f_\mathcal{N}(\mathbf{y}_i | \boldsymbol{\mu}_k, \boldsymbol{\Sigma}_k)$ in the subsequent classification step yields a non-negligible probability and, as a consequence, observations are more likely to be allocated to an empty component than in the standard prior case.

To illustrate the different allocation behavior of the standard and the normal gamma prior in the presence of a superfluous component more explicitly, we simulate $N = 1{,}000$ observations from a bivariate two-component mixture model where $\boldsymbol{\mu}_1 = (-2, 0)'$, $\boldsymbol{\mu}_2 = (2, 0)'$, $\boldsymbol{\Sigma}_1 = \boldsymbol{\Sigma}_2 = \mathbf{I}_2$, and $\boldsymbol{\eta} = (0.5, 0.5)$. We fit an overfitting mixture distribution with $K = 3$ components, assuming that $e_0 \sim \mathcal{G}(10, 30)$. We skip the random permutation step, since the modes of the posterior distribution are well separated and the sampler is trapped in the neighborhood of a single mode, yielding implicit identification.

In the top row of Fig. 2, posterior draws of all three component means are displayed in a scatter plot both for the standard (left-hand side) and the normal gamma prior (right-hand side). Under both priors, the posterior draws of the first two component means, displayed by triangle and cross points respectively, are concentrated around the true means $\boldsymbol{\mu}_1 = (-2, 0)'$ and $\boldsymbol{\mu}_2 = (2, 0)'$. However, the posterior draws of the mean of the third (superfluous) component, shown as circle points, are quite different, displaying a large dispersion over the plane under the standard prior and being located either close to the two true component means or the data center under the normal gamma prior. To illustrate the ensuing effect on classification, we select a specific observation $\mathbf{y}_i$, which location is marked by a (blue) star in the scatter plots of the top, and determine for each MCMC sweep the probability for $\mathbf{y}_i$ to be allocated, respectively, to component 1, 2 or 3. The corresponding box plots in the bottom row of Fig. 2 clearly indicate that the allocation probability for the third (superfluous) component is considerably higher under the normal gamma prior (plot on the right-hand side) than under the standard prior (plot on the left-hand side).

Since our strategy to estimate the number of mixture components relies on the number of non-empty components during MCMC sampling, we conclude from this investigation that stronger shrinkage in $\boldsymbol{\eta}$ might be necessary for the normal gamma prior (6) than for the standard prior (5). We compensate the tendency of the normal gamma prior to overestimate the number of non-empty components, by encouraging very small prior weights $\eta_k$ for empty components in order to keep the conditional probability of an observation to be allocated to an empty component during classification small. This is achieved by specifying a very small fixed hyperparameter $e_0$ in the Dirichlet prior, which is proportional to $\eta_k^{e_0-1}$. Thus, the smaller $e_0$, the smaller the weight of an empty component $k$ will be.

## 4 Identifying sparse finite mixtures

Identification of the finite mixture model requires handling the label switching problem caused by invariance of the representation (1) with respect to reordering the components:

$$f(\mathbf{y}_i | \boldsymbol{\theta}_1, \ldots, \boldsymbol{\theta}_K, \boldsymbol{\eta}) = \sum_{k=1}^{K} \eta_k f_k(\mathbf{y}_i | \boldsymbol{\theta}_k)$$
$$= \sum_{k=1}^{K} \eta_{\rho(k)} f_{\rho(k)}(\mathbf{y}_i | \boldsymbol{\theta}_{\rho(k)}),$$

where $\rho$ is an arbitrary permutation of $\{1, \ldots, K\}$. The resulting multimodality and perfect symmetry of the pos-terior distribution $p(\boldsymbol{\theta}_1, \ldots, \boldsymbol{\theta}_K, \boldsymbol{\eta}|\mathbf{y})$ for symmetric priors makes it difficult to perform component-specific inference. To solve the label switching problem arising in Bayesian mixture model estimation, it is necessary to post-process the MCMC output to obtain a unique labeling. Many useful methods have been developed to force a unique labeling on draws from this posterior distribution when the number of components is known (Celeux 1998; Celeux et al. 2000; Frühwirth-Schnatter 2001; Stephens 2000; Jasra et al. 2005; Yao and Lindsay 2009; Grün and Leisch 2009; Sperrin et al. 2010). However, most of these proposed relabeling methods become computationally prohibitive for multivariate data with increasing dimensionality. For instance, as explained in (Frühwirth-Schnatter 2006, p. 96), Celeux (1998) proposes to use a $K$-means cluster algorithm to allocate the draws of one iteration to one of $K!$ clusters, which initial centers are determined by the first 100 draws. The distance of the draws to each of the $K!$ reference centers is used to determine the labeling of the draws for this iteration. In general, most of the relabeling methods use the complete vector of parameters which grows as a multiple of $K$ even if they do not require all $K!$ modes to be considered (see, for example Yao and Lindsay 2009).

**Fig. 2** Fitting a 3-component normal mixture to data generated by a 2-component normal mixture. *Top row* Scatter plots of the draws of the posterior component means $\boldsymbol{\mu}_k$ under the standard prior (*left-hand side*) and the normal gamma prior (*right-hand side*). Draws from component 1, 2, and 3 are displayed as *green triangles*, *red crosses*, and *grey circles*, respectively. *Bottom row* For a single observation which location is marked with a (*blue*) *star* in the scatter plots in the *top row*, box plots of the conditional probabilities during MCMC sampling to be assigned to component 1, 2 or 3 are displayed, under the standard prior (*left-hand side*) and normal gamma prior (*right-hand side*). MCMC is run for 1,000 iterations, after discarding the first 1,000 draws. (Color figure online)

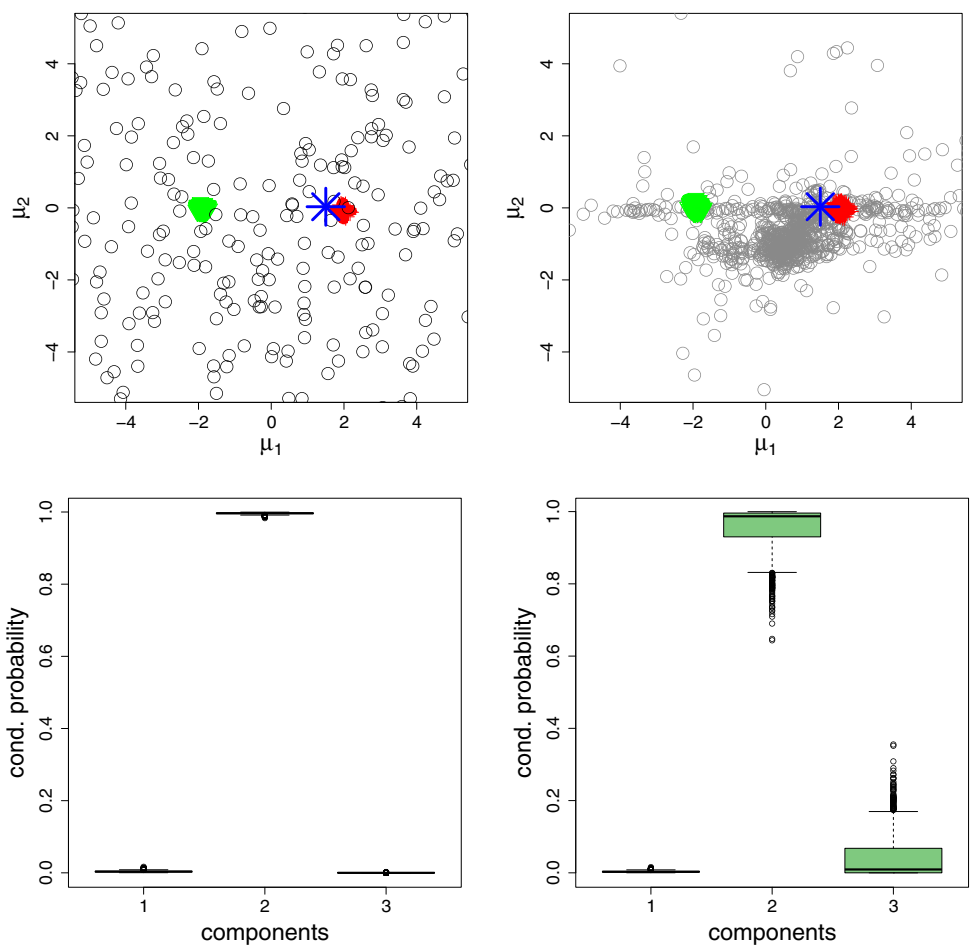

Following Frühwirth-Schnatter (2011a), we apply $K$-means clustering to the point process representation of the MCMC draws to identify a sparse finite mixture model, see Sect. 4.1. This allows to reduce the dimension of the problem to the dimension of the component-specific parameters. As described in Sect. 4.2, we generalize this approach by replacing $K$-means clustering by $K$-centroids cluster analysis based on the Mahalanobis distance (Leisch 2006).

### 4.1 Clustering the MCMC output in the point process representation

The point process representation of the MCMC draws introduced in Frühwirth-Schnatter (2006, Sect. 3.7.1) allows to study the posterior distribution of the component-specific parameters regardless of potential label switching, which makes it very useful for model identification. If the number of mixture components matches the true number of components, then the posterior draws of the component-specific parameters $\boldsymbol{\theta}_1^{(m)}, \ldots, \boldsymbol{\theta}_K^{(m)}$ cluster around the "true" points $\boldsymbol{\theta}_1^{true}, \ldots, \boldsymbol{\theta}_K^{true}$. To visualize the point process representation of the posterior mixture distribution, projections of the point process representation onto two-dimensional planes can be considered. These correspond to scatter plots of the MCMC draws $(\theta_{kj}^{(m)}, \theta_{kj'}^{(m)})$, for two dimensions $j$ and $j'$ and across $k = 1, \ldots, K$.

**Fig. 3** Crabs data, $K^{true} = 4$, $K = 15$, standard prior. Point process representation of posterior mean draws, $\mu_{k1}^{(m)}$ plotted against $\mu_{k2}^{(m)}$, across $k = 1, \ldots, K$. *Left-hand side* draws from *all* $K = 15$ components. *Right-hand side* only draws from those $M_0$ iterations where $K_0^{(m)} = 4$ and the components which were non-empty

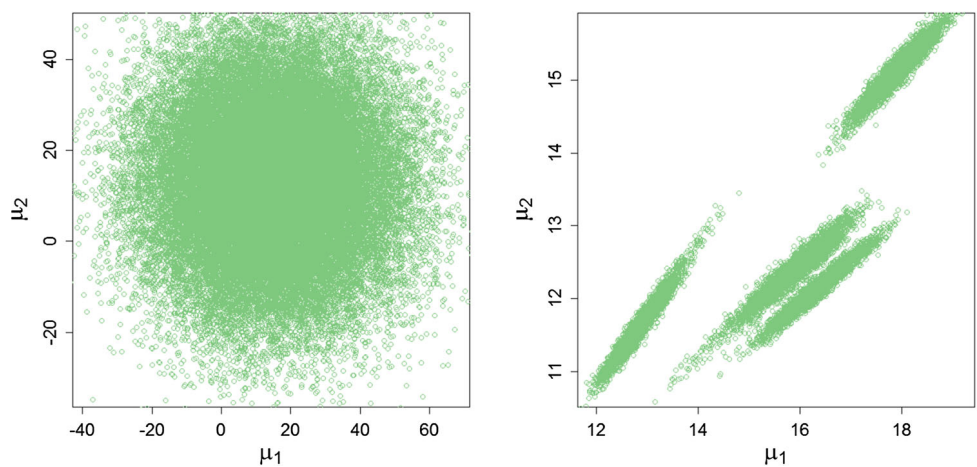

After clustering the draws in the point process representation, a unique labeling is achieved for all those draws where the resulting classification is a permutation. By reordering each of these draws according to its classification, a (subset) of identified draws is obtained which can be used for component-specific parameter inference.

Note that to reduce dimensionality, it is possible to cluster only a subset of parameters of the component-specific parameter vector and to apply the obtained classification sequences to the entire parameter vector. In the present context of multivariate Gaussian mixtures, we only clustered the posterior component means and the obtained classification sequence was then used to reorder and identify the other component-specific parameters, namely covariance matrices and weights. This is also based on the assumption that the obtained clusters differ in the component means allowing clusters to be characterized by their means.

In the case of a sparse finite mixture model, where the prior on the weights favors small weights, many of the components will have small weights and no observation will be assigned to these components in the classification step. Component means of all empty components are sampled from the prior and tend to confuse the cluster fit. Therefore, Frühwirth-Schnatter (2011a) suggests to remove all draws from empty components before clustering. Additionally, after having estimated the number of non-empty components $\hat{K}_0$, all draws where the number of non-empty components is different from $\hat{K}_0$ are sampled conditional on a "wrong" model and are removed as well. The remaining draws can be seen as samples from a mixture model with exactly $\hat{K}_0$ components. In Fig. 3 an example of the point process representation of the MCMC draws is given. After having fitted a sparse finite mixture with $K = 15$ components to the Crabs data set described in Sect. 5.2, the left-hand side shows the scatter plot of the MCMC draws $(\mu_{k1}^{(m)}, \mu_{k2}^{(m)})$, $k = 1, \ldots, K$, from *all* components (including draws from empty components). On the right-hand side, only draws from those $M_0$ iterations are plotted where $\hat{K}_0 = 4$ and which were non-empty. In this case, the posterior distributions of the means of the four non-empty components can be clearly distinguished. These draws are now clustered into $\hat{K}_0$ groups. The clusters naturally can be assumed to be of equal size and to have an approximate Gaussian distribution, thus suggesting the choice of $K$-means for clustering or, in order to capture also non-spherical shapes or different volumes of the posterior distributions, the choice of $K$-centroids cluster analysis where the distance is defined by a cluster-specific Mahalanobis distance. The algorithm is explained in the following subsection. The detailed scheme to identify a sparse finite mixture model can be found in Appendix 2.

### 4.2 $K$-centroids clustering based on the Mahalanobis distance

Defining the distance between a point and a centroid using the Mahalanobis distance may considerably improve the cluster fit in the point process representation. As can be seen in Fig. 4, where the clustering results for the Crabs data are displayed, if the posterior distributions have elliptical shape, clustering based on the Mahalanobis distance is able to catch the elongated, elliptical clusters whereas $K$-means based on the squared Euclidean distance splits a single cluster into several parts and at the same time combines these parts to one new artificial cluster.

For posterior draws of the component-specific parameter vector $\mathbf{x}_1, \ldots, \mathbf{x}_N \in \mathbb{R}^n$ and a fixed number of clusters $K$, the $K$-centroids cluster problem based on the Mahalanobis distance consists of finding a "good" set of centroids and dispersion matrices

$$CS_K = \{\mathbf{c}_1, \ldots, \mathbf{c}_K, \mathbf{S}_1, \ldots, \mathbf{S}_K\}, \tag{11}$$

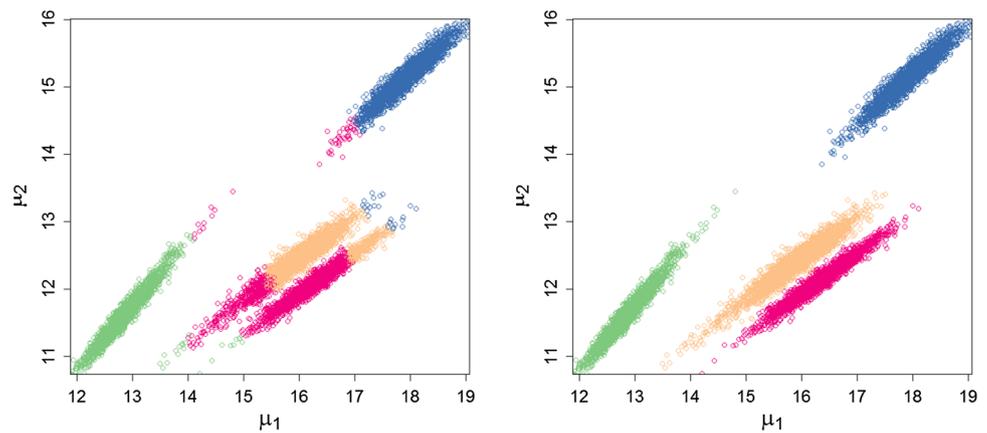

where $\mathbf{c}_1, \ldots, \mathbf{c}_K$ are points in $\mathbb{R}^n$ and $\mathbf{S}_1, \ldots, \mathbf{S}_K$ are instances of the set of all positive definite matrices. "Good" means that using the assigned dispersion matrices $\mathbf{S}(\mathbf{x}_i)$, the sum of all distances between objects $\mathbf{x}_i$ and their assigned centroids $\mathbf{c}(\mathbf{x}_i)$ is minimized:

$$\sum_{i=1}^{N} d_{\mathbf{S}(\mathbf{x}_i)}(\mathbf{x}_i, \mathbf{c}(\mathbf{x}_i)) \to \min_{\mathbf{c}_1, \ldots, \mathbf{c}_K, \mathbf{S}_1, \ldots, \mathbf{S}_K}, \qquad (12)$$

where $\{\mathbf{c}(\mathbf{x}_i), \mathbf{S}(\mathbf{x}_i)\} = \operatorname*{argmin}_{\{\mathbf{c}_k, \mathbf{S}_k\} \in CS_K} d_{\mathbf{S}_k}(\mathbf{x}_i, \mathbf{c}_k),$

and the distance between an object $\mathbf{x}_i$ and a centroid and a dispersion matrix $(\mathbf{c}_k, \mathbf{S}_k)$ is defined by the Mahalanobis distance:

$$d_{\mathbf{S}_k}(\mathbf{x}_i, \mathbf{c}_k) = \sqrt{(\mathbf{x}_i - \mathbf{c}_k)' \mathbf{S}_k^{-1} (\mathbf{x}_i - \mathbf{c}_k)}. \qquad (13)$$

Since no closed form solution exists for the $K$-centroids cluster problem, an iterative estimation procedure is used. A popular choice is the well-known $K$-means algorithm, its general form can be found in Leisch (2006). For the Mahalanobis distance (13), the $K$-centroids cluster algorithm is given by:

1. Start with a random set of initial centroids and variance-covariance matrices $\{\mathbf{c}_k, \mathbf{S}_k\}_{k=1,\ldots,K}$.
2. Assign to each $\mathbf{x}_i$ the nearest centroid $\mathbf{c}_k$ where the distance $d_{\mathbf{S}_k}(\mathbf{x}_i, \mathbf{c}_k)$ is defined by the Mahalanobis distance (13):

   $$\{\mathbf{c}(\mathbf{x}_i), \mathbf{S}(\mathbf{x}_i)\} = \operatorname*{argmin}_{\{\mathbf{c}_k, \mathbf{S}_k\} \in CS_K} d_{\mathbf{S}_k}(\mathbf{x}_i, \mathbf{c}_k)$$

3. Update the set of centroids and dispersion matrices $\{\mathbf{c}_k, \mathbf{S}_k\}_{k=1,\ldots,K}$ holding the cluster membership fixed:

   $$\mathbf{c}_k^{(new)} = \operatorname*{mean}_{i:\mathbf{c}(\mathbf{x}_i)=\mathbf{c}_k, \mathbf{S}(\mathbf{x}_i)=\mathbf{S}_k}(\{\mathbf{x}_i\}),$$
   $$\mathbf{S}_k^{(new)} = \operatorname*{var}_{i:\mathbf{c}(\mathbf{x}_i)=\mathbf{c}_k, \mathbf{S}(\mathbf{x}_i)=\mathbf{S}_k}(\{\mathbf{x}_i\})$$

4. Repeat steps 2 and 3 until convergence.

The algorithm is guaranteed to converge in a finite number of iterations to a local optimum of the objective function (12) (Leisch 2006). As starting values of the algorithm in step 1, the MAP estimates of the hyperparameters $\mathbf{b}_1, \ldots, \mathbf{b}_K, \mathbf{B}_1, \ldots, \mathbf{B}_K$ of the prior of the component-specific means are used.

# 5 Simulations and applications

## 5.1 Simulation study

In the following simulation study, the performance of the proposed strategy for selecting the unknown number of mixture components and identifying cluster-relevant variables is illustrated for the case where the component densities are truly multivariate Gaussian. We use a mixture of four multivariate Gaussian distributions with component means $\boldsymbol{\mu}_1 = (2, -2, 0, 0)'$, $\boldsymbol{\mu}_2 = -\boldsymbol{\mu}_1$, $\boldsymbol{\mu}_3 = (2, 2, 0, 0)'$, and $\boldsymbol{\mu}_4 = -\boldsymbol{\mu}_3$ and isotropic covariance matrices $\boldsymbol{\Sigma}_k = \mathbf{I}_4$, $k = 1, \ldots, 4$, as data generating mechanism. Hence, this simulation setup consists of two cluster-generating variables in dimension 1 and 2 and two homogeneous variables in dimension 3 and 4 and is chosen in order to study, whether cluster-relevant variables and homogeneous variables can be distinguished. In Fig. 5, a randomly selected data set is shown, which was sampled with equal weights. This figure indicates that, while in the scatter plot of the first two variables four clusters are still visually discernible, the clusters are slightly overlapping in these dimensions indicating that the cluster membership of some observations is difficult to estimate. The other two variables are homogenous variables and do not provide any cluster information, but render the clustering task more difficult.

As described in Sect. 2.1, we deliberately choose an overfitting mixture with $K$ components and base our estimate of the true number of mixture components on the frequency of non-empty components during MCMC sampling. We select strongly overfitting mixtures with $K = 15$ and $K = 30$, to assess robustness of the proposed strategy to choosing $K$, and investigate, if the number of non-empty components increases as $K$ increases. We simulate relatively large samples of 1,000 observations to make it more difficult to really empty all superfluous components.

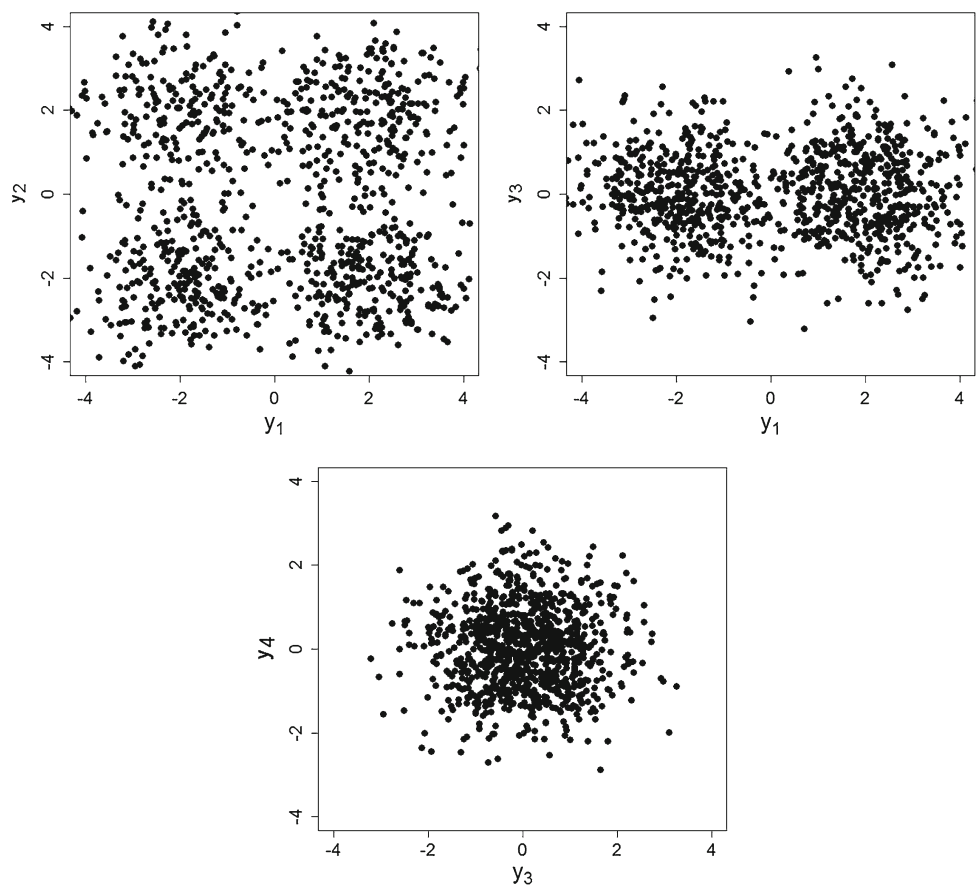

In addition, we consider two different weight distributions, namely a mixture with equal weights, i.e. $\boldsymbol{\eta} = (0.25, 0.25, 0.25, 0.25)$, and a mixture with a very small component, i.e. $\boldsymbol{\eta} = (0.02, 0.33, 0.33, 0.32)$, in order to study how sensitive our method is to different cluster sizes. For the second mixture, we investigate whether the small component survives or is emptied during MCMC sampling together with all superfluous components.

Prior distributions and the corresponding hyperparameters are specified as described in Sect. 2. The prior on the weight distribution defines a sparse finite mixture distribution. We either use the gamma prior on $e_0$ defined in (3) or choose a very small, but fixed value for $e_0$ as motivated by Sect. 3.2. In addition, we compare the standard prior (5) for the component means with the hierarchical shrinkage prior (6) based on the normal gamma distribution.

For each setting, ten data sets are generated, each consisting of 1,000 data points $\mathbf{y}_i$, and MCMC sampling is run for each data set for $M = 10,000$ iterations after a burn-in of

2,000 draws. The starting classification of the observations is obtained by $K$-means. Estimation results are averaged over the ten data sets and are reported in Tables 1 and 2 where $\hat{e}_0$ provides the posterior median of $e_0$ under the prior (3), whereas "$e_0$ fixed" indicates that $e_0$ was fixed at the reported value. $\hat{K}_0$ is the posterior mode estimator of the true number of non-empty components which is equal to 4. If the estimator $\hat{K}_0$ did not yield the same value for all data sets, then the number of data sets where $\hat{K}_0$ was the estimated number of non-empty components is given in parentheses. $M_0$ is the average number of iterations where exactly $\hat{K}_0$ components were non-empty.

For each data set, these draws are identified as described in Sect. 4 using clustering in the point process representation. $M_{0,\rho}$ is the (average) rate among the $M_0$ iterations where the corresponding classifications assigned to the draws by the clustering procedure fail to be a permutation. Since in these cases the labels $1, \ldots, \hat{K}_0$ cannot be assigned uniquely to the $\hat{K}_0$ components, these draws are discarded. For illustration, see the example in the Appendix 2. The non-permutation rate $M_{0,\rho}$ is a measure for how well-separated the posterior distributions of the component-specific parameters are in the point process representation, with a value of 0 indicating perfect separation, see Appendix 2 and Frühwirth-Schnatter (2011a) for more details. In the following component-specific inference is based on the remaining draws where a unique labeling was achieved.

**Table 1** Simulation setup with equal weights: results for different $K$ under the standard prior (Sta), the normal gamma Prior (Ng), and when fitting an infinite mixture using the R package PReMiuM. Results are averaged over ten data sets

| Prior | $K$ | $\hat{e}_0$ | $e_0$ fixed | $\hat{K}_0$ | $M_0$ | $M_{0,\rho}$ | $MCR$ | $MSE_\mu$ |
|---|---|---|---|---|---|---|---|---|
| Sta | 4 | 0.28 | | 4 | 10000 | 0 | 0.049 | 0.167 |
| | 15 | 0.05 | | 4 | 9802 | 0 | 0.049 | 0.167 |
| | 30 | 0.03 | | 4 | 9742 | 0 | 0.048 | 0.168 |
| Ng | 4 | 0.28 | | 4 | 10000 | 0 | 0.049 | 0.137 |
| | 15 | 0.06 | | **5**(6) | 2845 | 0.85 | 0.050 | — |
| | 30 | 0.03 | | **5**(5) | 2541 | 0.93 | 0.050 | — |
| Ng | 4 | | 0.01 | 4 | 10000 | 0 | 0.047 | 0.136 |
| | 15 | | 0.01 | 4 | 7465 | 0 | 0.048 | 0.137 |
| | 30 | | 0.01 | **4**(8) | 4971 | 0 | 0.048 | 0.136 |
| | 30 | | 0.001 | 4 | 9368 | 0 | 0.048 | 0.136 |
| | 30 | | 0.00001 | 4 | 9998 | 0 | 0.047 | 0.136 |
| PReMiuM | | $\hat{\alpha}$ | | $K^{est}$ | | | $MCR$ | $MSE_\mu$ |
| | | 0.66 | | 4 | | | 0.047 | 0.231 |

**Table 2** Simulation setup with unequal weights: results for different $K$ under the standard prior (Sta), the normal gamma prior (Ng), and when fitting an infinite mixture using the R package PReMiuM. Results are averaged over ten data sets

| Prior | $K$ | $\hat{e}_0$ | $e_0$ fixed | $\hat{K}_0$ | $M_0$ | $M_{0,\rho}$ | $MCR$ | $MSE_\mu$ |
|---|---|---|---|---|---|---|---|---|
| Sta | 4 | 0.27 | | 4 | 10000 | 0.00 | 0.038 | 1.670 |
| | 15 | 0.05 | | 4 | 9780 | 0.00 | 0.037 | 1.668 |
| | 30 | 0.03 | | 4 | 9728 | 0.00 | 0.038 | 1.663 |
| Ng | 4 | | 0.01 | 4 | 10000 | 0.02 | 0.037 | 1.385 |
| | 15 | | 0.01 | 4 | 7517 | 0.02 | 0.038 | 1.314 |
| | 30 | | 0.01 | **4**(9) | 5221 | 0.00 | 0.037 | 1.279 |
| | 30 | | 0.001 | 4 | 9297 | 0.01 | 0.037 | 1.325 |
| | 30 | | 0.00001 | 4 | 9997 | 0.02 | 0.038 | 1.336 |
| PReMiuM | | $\hat{\alpha}$ | | $K^{est}$ | | | $MCR$ | $MSE_\mu$ |
| | | 0.65 | | **4**(9) | | | 0.038 | 2.841 |

Accuracy of the estimated mixture model is measured by two additional criteria. Firstly, we consider the misclassification rate ($MCR$) of the estimated classification. The estimated classification of the observations is obtained by assigning each observation to the component where it has been assigned to most often during MCMC sampling among the draws where $\hat{K}_0$ components were non-empty and which could be uniquely relabeled. The misclassification rate is measured by the number of misclassified observations divided by all observations and should be as small as possible. The labeling of the estimated classification is calculated by "optimally" matching true compo-nents to the estimated mixture components obtaining in this way the minimum misclassification rate over all possible matches.

Secondly, whenever the true number of mixture components is selected for a data set, the mean squared error of the estimated mixture component means ($MSE_\mu$) based on the Mahalanobis distance is determined as

$$MSE_\mu = \sum_{k=1}^{\hat{K}_0} \frac{1}{\tilde{M}_0} \sum_{m=1}^{\tilde{M}_0} (\boldsymbol{\mu}_k^{(m)} - \boldsymbol{\mu}_k^{true})' (\boldsymbol{\Sigma}_k^{true})^{-1} (\boldsymbol{\mu}_k^{(m)} - \boldsymbol{\mu}_k^{true}),$$

(14)

where $\tilde{M}_0 = M_0(1 - M_{0,\rho})$ is the number of identified draws with exactly $\hat{K}_0$ non-empty components. In Sects. 5.2 and 5.3, where the true parameters $\boldsymbol{\mu}_k$ and $\boldsymbol{\Sigma}_k$ are unknown, the

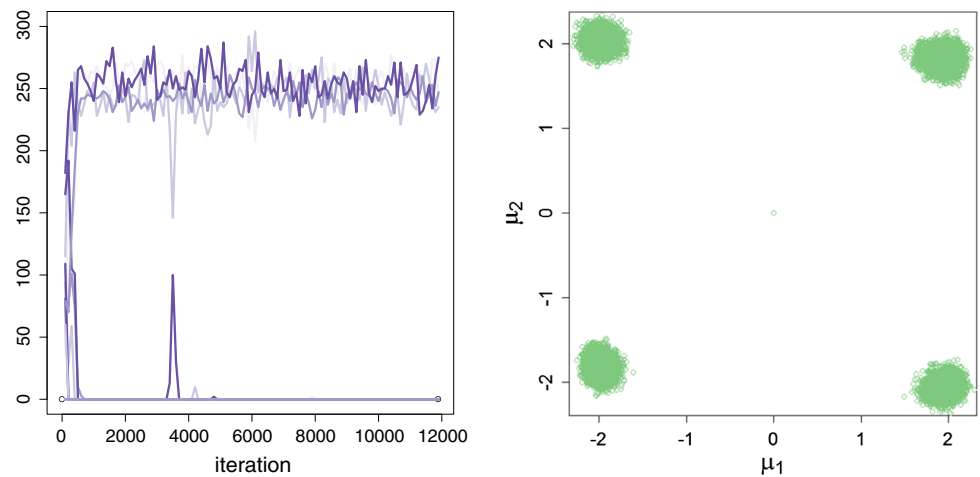

Bayes estimates of the parameters are taken instead. They are calculated by running the MCMC sampler with known allocations and taking the mean of the corresponding posterior draws. Evidently, $MSE_\mu$ should be as small as possible.

For comparison, results are also reported for a finite mixture, where $K = 4$ is known to be equal to the true value.

Finally, to compare our results to the clustering results obtained by fitting infinite mixtures, the R package PReMiuM (Liverani et al. 2013) is used to compute a Dirichlet process mixture model $DP(\alpha)$ with concentration parameter $\alpha$. The number of initial clusters is set to 30, the number of burn-in draws and number of iterations are set to 2,000 and 10,000, respectively. All other settings, such as hyperparameter specifications, calculation of the similarity matrix and the derivation of the best partition, are left at the default values of the package. After having obtained the estimated number of groups $K^{est}$ and the best partition vector $z^{best}$, in a post-processing way identification and summarization methods are applied on the MCMC output as proposed in Molitor et al. (2010). To obtain the posterior distributions of the component means of each group, for each iteration $m$ the average value $\bar{\boldsymbol{\mu}}_k^{(m)}$ of group $k$ is computed by:

$$\bar{\boldsymbol{\mu}}_k^{(m)} = \frac{1}{N_k} \sum_{i\,:\,z_i^{best} = k} \boldsymbol{\mu}_{S_i^{(m)}}^{(m)}, \tag{15}$$

where $N_k$ denotes the number of individuals in group $k$ of $z^{best}$ and $S_i^{(m)}$ is the component to which observation $i$ was assigned in iteration $m$. By averaging over all associated cluster means in each iteration, the posterior distribution of the component means is reconstructed in a post-processing way and represents the uncertainty associated with the final classification of the observations. Therefore, a larger MSE is expected than for our approach where the assumed model, a finite mixture of Gaussian distributions, corresponds to the true underlying model and the posterior component mean distribution is directly estimated during MCMC sampling.

The $MSE_\mu$ is then computed using Eq. (14). In the tables, the posterior mean $\hat{\alpha}$ of the concentration parameter $\alpha$ is reported. Following Ishwaran et al. (2001), the estimated $\alpha$ can be compared to $e_0$ using that a sparse finite mixture model with prior on the weights $Dir(e_0, \ldots, e_K)$ approximates an infinite mixture model with Dirichlet process prior $DP(\alpha)$ if $e_0 = \alpha/K$, see Sect. 2.1.

### 5.1.1 Simulation setup with equal weights

Table 1 reports the results for the simulation setup with equal weights. Under the standard prior, the estimated number of non-empty mixture components $\hat{K}_0$ matches the true number of components for both overfitting mixtures for all data sets, regardless whether $K = 15$ or $K = 30$ components have been used. Furthermore, exactly four components are non-empty for most of the draws, since $M_0 \approx M$. The non-permutation rate $M_{0,\rho}$ is zero for all overfitting mixtures, indicating that the posterior distributions of all $\hat{K}_0$ non-empty components are well-separated in the point process representation.

MCMC estimation for an overfitting mixture with $K = 15$ components is explored in more detail in Fig. 6 for a randomly selected data set. The trace plot of allocations displays the number of observations allocated to the different components during MCMC sampling, also the burn-in draws are considered. Due to the starting classification through $K$-means, the observations are assigned to all components at the beginning, however, rather quickly all but four components are emptied. In the scatter plot of the point process representation of the component mean draws of non-empty components $(\mu_{k1}^{(m)}, \mu_{k2}^{(m)})$, $k = 1, \ldots, K$, sampled from exactly $\hat{K}_0 = 4$ non-empty components, it can be seen very clearly that the draws cluster around the true parameter values $(2, 2)$, $(2, -2)$, $(-2, 2)$ and $(-2, -2)$.

If $e_0$ is considered to be random with prior (3), the estimated Dirichlet parameter $\hat{e}_0$ has a very small value, much smaller than the (asymptotic) upper bound given by Rousseau and Mengersen (2011), and decreases, as the number of redundant components increases. This is true both for the standard prior and the normal gamma prior. However, under the normal gamma prior, the estimated number of non-empty components $\hat{K}_0$ overfits the true number of mixture components for most of the data sets and increases with $K$. For example, if $K = 15$, the number of non-empty components is overestimated with $\hat{K}_0 = 5$ for 6 data sets. Also the high average non-permutation rate $M_{0,\rho} = 0.85$ indicates that the selected model is overfitting. However, the $MCR$ is not higher than for $K = 4$, indicating that the fifth non-empty component is a very small one.

Given the considerations in Sect. 3.2, we combine the normal gamma prior with a sparse prior on the weight distribution where $e_0$ is set to a fixed very small value, e.g. 0.01 which is smaller than the 1 % quantile of the posterior of $e_0$. For this combination of priors, superfluous components are emptied also under the normal gamma prior and the estimated number of non-empty components matches the true number of mixture components in most cases. For $K = 30$, $e_0$ has to be chosen even smaller, namely equal to 0.001, to empty *all* superfluous components for all data sets. To investigate the effect of an even smaller value of $e_0$, we also set $e_0 = 10^{-5}$. Again, four groups are estimated. Thus evidently as long as the cluster information is strong small values of $e_0$ do not lead to underestimating the number of clusters in a data set. As a consequence, for the following simulations, we generally combine the normal gamma distribution with a sparse prior on the weight distribution where $e_0$ is always set to fixed, very small values.

Both for the standard prior (with $e_0$ random) and the normal gamma prior (with $e_0$ fixed), the misclassification rate $MCR$ and the mean-squared error $MSE_\mu$ of the estimated models have the same size, as if we had known the number of mixture components in advance to be equal to $K = 4$. This oracle property of the sparse finite mixture approach is very encouraging.

While the misclassification rate $MCR$ is about the same for both priors, interestingly, the $MSE_\mu$ is considerably smaller under the normal gamma prior ($\approx 0.136$) than under the standard prior ($\approx 0.167$) for all $K$. This gain in efficiency illustrates the merit of choosing a shrinkage prior on the component-specific means.

As noted in Sect. 2.2, a further advantage of specifying a normal gamma prior for the component means, is the possibility to explore the posterior distribution of the scaling factor $\lambda_j$. Therefore, visual inspection of the box plots of the posterior draws of $\lambda_j$ helps to distinguish between variables, where component distributions are well separated, and variables, where component densities are strongly overlapping or even identical. The box plots of the posterior draws of $\lambda_j$ displayed in Fig. 8 clearly indicate that only the first two variables show a high dispersion of the component means, whereas for the two other dimensions the posterior distribution of $\lambda_j$ is strongly pulled toward zero indicating that component means are located close to each other and concentrate at the same point.

If the data sets are clustered by fitting an infinite mixture model with the R package PReMiuM, similar clustering results are obtained. For all ten data sets four groups are estimated. The averaged estimated concentration parameter $\hat{\alpha}$ is 0.66. This indicates, that a sparse finite mixture model with $K = 30$ components and $e_0 \approx 0.02$ is a good approximation to a Dirichlet process DP($\alpha$) as $\alpha/K = 0.66/30 = 0.022$. As expected, the $MSE_\mu$ of the cluster means is considerable larger (0.231) than for sparse finite mixtures, whereas the misclassification rate (0.047) is as for finite mixtures.

### 5.1.2 Simulation setup with unequal weights

To study if small non-empty components can be identified under a sparse prior on the weights, the second simulation setup uses the weight distribution $\boldsymbol{\eta} = (0.02, 0.33, 0.33, 0.32)$ for data generation, where the first component generates only 2 % of the data.

The simulation results are reported in Table 2. Regardless of the number of specified components $K$, $\hat{K}_0 = 4$ non-empty components are identified under both priors. Again, for the normal gamma prior, the hyperparameter $e_0$ of the Dirichlet distribution has to be set to a very small value (0.001 or even smaller) to empty all superfluous components in all data sets.

While our estimator $\hat{K}_0$ is robust to the presence of a very small component, selecting the number of components by identifying "large" weights, as has been suggested by Rousseau and Mengersen (2011), is likely to miss the fourth small component. In the left-hand side plot of Fig. 7, the (unidentified) posterior weights sorted by size in each iteration are displayed for a single data set. The forth largest weight in each iteration is very small and there might be uncertainty whether the forth component belongs to either the superfluous components or constitutes a non-empty component. However, by looking for non-empty components during MCMC sampling as our approach does, the small component can be clearly identified since it is never emptied during the whole MCMC run, as can be seen in the trace plot of allocations in Fig. 7.

Again, both for the standard prior and the normal gamma prior, the misclassification rate $MCR$ and the mean-squared error $MSE_\mu$ of the estimated models have the same size, as if we had known the number of mixture components in advance to be equal to $K = 4$. Again, the normal gamma prior dominates the standard prior, with the $MSE_\mu$ being considerably smaller under the normal gamma prior ($\approx 1.385$) than under the standard prior ($\approx 1.670$) for all $K$. This illustrates once

**Fig. 7** Simulation setup with unequal weights, diagnostic plots of a single MCMC run, $K = 30$, standard prior: Box plots of the (unidentified) posterior weight draws, sorted by size in each iteration (*left-hand side*) and trace plot of the number of observations allocated to the different mixture components, burn-in included (*right-hand side*)

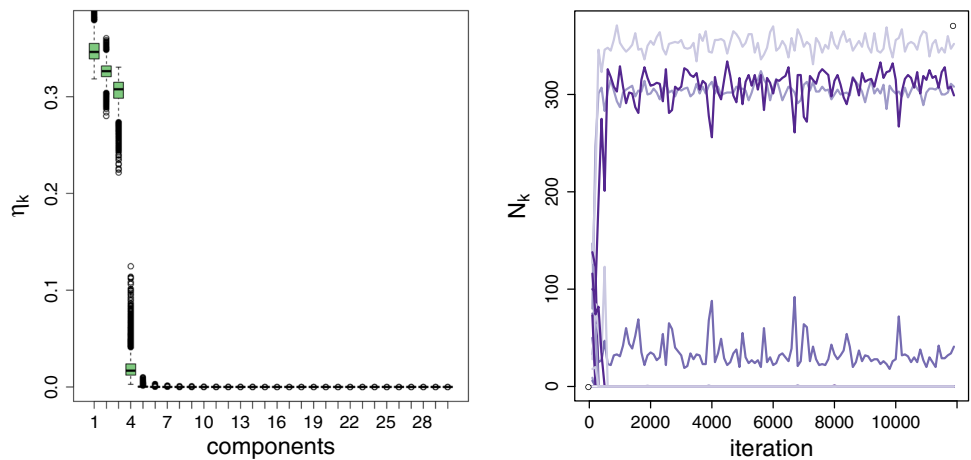

**Fig. 8** $K = 15$, normal gamma prior: Box plots of shrinkage factors $\lambda_j$, for the simulation setup with equal weights (*left-hand side*) and unequal weights (*right-hand side*). Posterior draws of *all* data sets are plotted

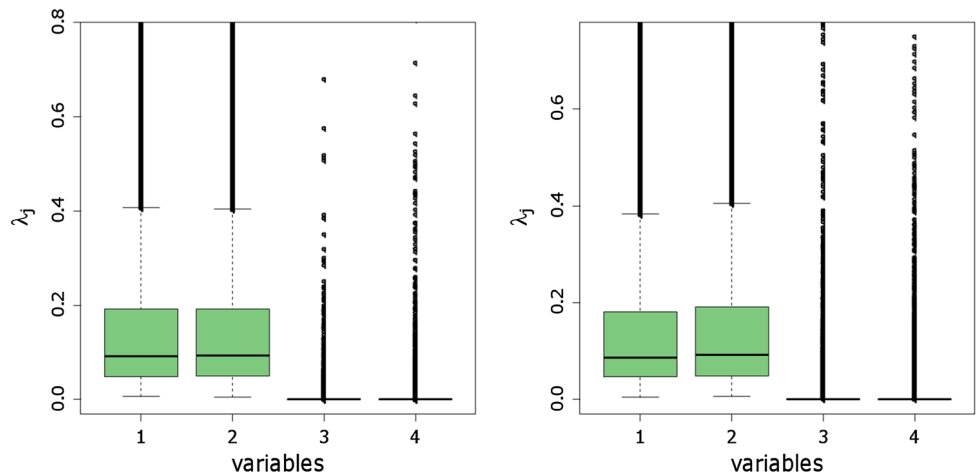

more the efficiency gain of choosing a shrinkage prior on the component-specific means.

Also for this simulation setting, the posterior distribution of scaling factors $\lambda_j$, sampled for $K = 15$ under the normal gamma prior and displayed in Fig. 8 on the right-hand side, clearly indicates that only the first two variables are cluster-generating, regardless of the small cluster size of the first component.

Again similar clustering results are obtained when fitting infinite mixtures. For almost all data sets (9 out of 10) the true number of groups is estimated. Again the $MSE_\mu$ is larger for infinite mixtures (2.841) than for sparse finite mixtures.

### 5.2 Crabs data

The Crabs data set, first published by Campbell and Mahon (1974) and included in the R package MASS (Venables and Ripley 2002), consists of 200 observations of a crabs population. It describes five morphological measurements on 200 crabs which have one of two color forms and differ in sex. Thus, four different groups are "pre-defined" and in the following we aim at recovering these four groups using the sparse finite mixture approach. Thus we would expect to find four data clusters. However, the correct number of clusters may be more than four (if some of the "pre-defined" groups are heterogeneous themselves) or less than four (if some of the "pre-defined" groups are not distinguishable), see considerations made by Hennig (2010). Among others, the data set was analyzed by Raftery and Dean (2006), Hennig (2010) and Yau and Holmes (2011). We used the original data without transformations.

For selecting the number of mixture components, sparse finite mixture models with $K = 15$ and $K = 30$ mixture components are specified. As can be seen in Table 3, under both the standard and the normal gamma prior the expected number of components $\hat{K}_0 = 4$ is selected. The posterior distribution converges rather fast to four non-empty components, as can be seen in the trace plot on the left-hand side in Fig. 9, where the number of observations allocated to the 15 components are displayed.

The misclassification rate $MCR$ of the identified model is 0.08 for the standard prior and 0.07 for the normal gamma prior. In Raftery and Dean (2006) the misclassification rate was 0.40 when using all variables as we do, and 0.075 when

 

**Table 3** Crabs data: results for different $K$ under the standard prior (Sta) and the normal gamma prior (Ng), and when fitting an infinite mixture using the R package PReMiuM. The $MSE_\mu$ is calculated using the Mahalanobis distance based on Bayes estimates. $M'_{0,\rho}$, $MCR'$, and $MSE'_\mu$ are the results based on the clustering of the draws in the point process representation through $K$-means instead of the $K$-centroids cluster analysis based on the Mahalanobis distance

| Prior | $K$ | $\hat{e}_0$ | $e_0$ fixed | $\hat{\mathbf{K}}_0$ | $M_0$ | $M_{0,\rho}$ | $MCR$ | $MSE_\mu$ | $M'_{0,\rho}$ | $MCR'$ | $MSE'_\mu$ |
|---|---|---|---|---|---|---|---|---|---|---|---|
| Sta | 4 | 0.27 | | **4** | 10,000 | 0 | 0.08 | 0.80 | 0.27 | 0.08 | 3.67 |
| | 15 | 0.05 | | **4** | 10,000 | 0 | 0.08 | 0.81 | 0.28 | 0.08 | 3.82 |
| | 30 | 0.03 | | **4** | 10,000 | 0 | 0.08 | 0.80 | 0.29 | 0.08 | 3.42 |
| Ng | 4 | | 0.01 | **4** | 10,000 | 0 | 0.07 | 0.68 | 0.44 | 0.08 | 6.72 |
| | 15 | | 0.01 | **4** | 9,938 | 0 | 0.07 | 0.67 | 0.46 | 0.08 | 8.19 |
| | 30 | | 0.01 | **4** | 9,628 | 0 | 0.07 | 0.68 | 0.46 | 0.08 | 8.10 |
| PReMiuM | | $\hat{\alpha}$ | | $K^{est}$ | | | $MCR$ | $MSE_\mu$ | | | |
| | | 0.67 | | **3** | | | 0.28 | | | | |

**Fig. 9** Crabs data, normal gamma prior, $K = 15$: Trace plot of the number of observations allocated to the different components, burn-in included (*left-hand side*). Box plots of the shrinkage factors $\lambda_j$ for all five variables (*right-hand side*)

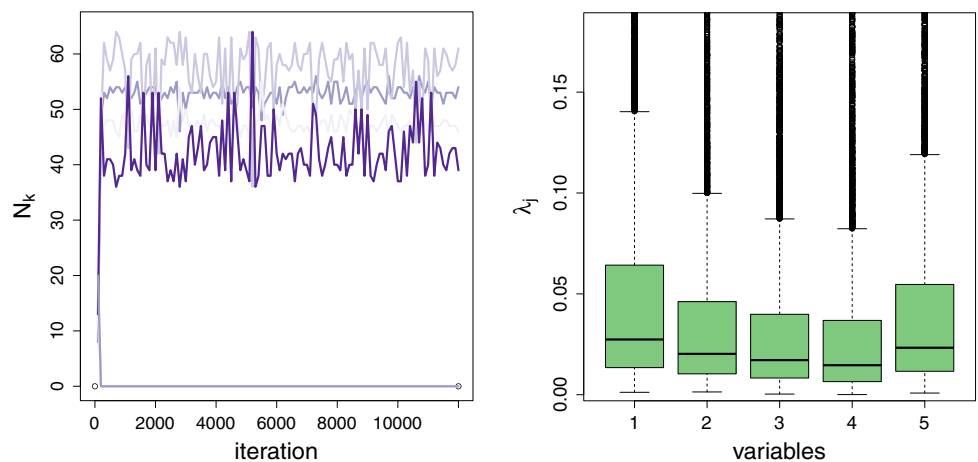

excluding one variable. Again, there is a considerable reduction in $MSE_\mu$ under the normal gamma prior compared to the standard prior. Box plots of the posterior draws of the shrinkage factor $\lambda_j$ in Fig. 9 reveal that all five variables are cluster-relevant which might be due to the high correlation between variables.

This specific case study also demonstrates the importance of the refined procedure introduced in Sect. 4.2 to identify a mixture by clustering the MCMC draws of the component-specific means in the point process representation. Clustering using the squared Euclidean distance fails to capture the geometry of the posterior mean distribution and leads to a high non-permutation rate, denoted by $M'_{0,\rho}$ in Table 3. Clustering using $K$-centroids cluster analysis based on the Mahalanobis distance, however, allows to capture the elliptical shapes of the posterior mean distribution properly, see Fig. 4, which in turn reduces the non-permutation rate $M_{0,\rho}$ to 0. In this way, inference with respect to the component-specific parameters is considerably improved, as is evident from comparing $MSE_\mu$ and $MSE'_\mu$ for both clustering methods in Table 3.

By clustering the Crabs data using an infinite mixture model with initial settings as explained in Sect. 5.1, three groups are estimated.

### 5.3 Iris data

The Iris data set (Anderson 1935; Fisher 1936) consists of 50 samples from each of three species of Iris, namely Iris setosa, Iris virginica and Iris versicolor. Four features are measured for each sample, the length and width of the sepals and petals, respectively. We aim at recovering the three underlying classes using the sparse finite mixture approach and thus expect to find three data clusters, although, as mentioned already for the Crabs data in Sect. 5.2, the true number of clusters for a finite mixture of Gaussian distributions may be more or less than three.

The results are reported in Table 4 by fitting sparse finite mixture models with 15 and 30 components, respectively. Values given in parentheses refer to the draws associated with the number of non-empty components given in parenthesis in column $\hat{K}_0$. Under both priors, the expected number of com-ponents is selected, as the majority of the draws is sampled from a mixture model with exactly three non-empty components. Under the standard prior, the misclassification rate of the identified model is 0.027, which outperforms the rate of 0.04 given in Raftery and Dean (2006).

**Table 4** Iris data: results for different $K$, under the standard prior (Sta) and the normal gamma prior (Ng), and when fitting an infinite mixture using the R package PReMiuM. The $MSE_\mu$ is calculated using the Mahalanobis distance based on Bayes estimates. Values given in parentheses refer to the draws associated with the number of non-empty components given in parenthesis in column $\hat{K}_0$

| Prior | $K$ | $\hat{e}_0$ | $e_0$ fixed | $\hat{K}_0$ | $M_0$ | $M_{0,\rho}$ | $MCR$ | $MSE_\mu$ |
|---|---|---|---|---|---|---|---|---|
| Sta | 3 | 0.34 | | **3** | 10,000 | 0 | 0.027 | 0.338 |
| | 15 | 0.05 | | **3** (4) | 5,900 (4086) | 0 (0.004) | 0.027 (0.093) | 0.336 |
| | 30 | 0.03 | | **3** (4) | 6889 (3111) | 0 (0.002) | 0.027 (0.093) | 0.338 |
| Ng | 3 | | 0.01 | **3** | 10,000 | 0 | 0.027 | 0.350 |
| | 15 | | 0.01 | **3** (4) | 7469 (2496) | 0 (0.043) | 0.033 (0.093) | 0.343 |
| | 30 | | 0.01 | **3** (4) | 6157 (3730) | 0 (0.147) | 0.033 (0.093) | 0.349 |
| PReMiuM | | $\hat{\alpha}$ | | $K^{est}$ | | | $MCR$ | $MSE_\mu$ |
| | | 0.52 | | **2** | | | 0.33 | |

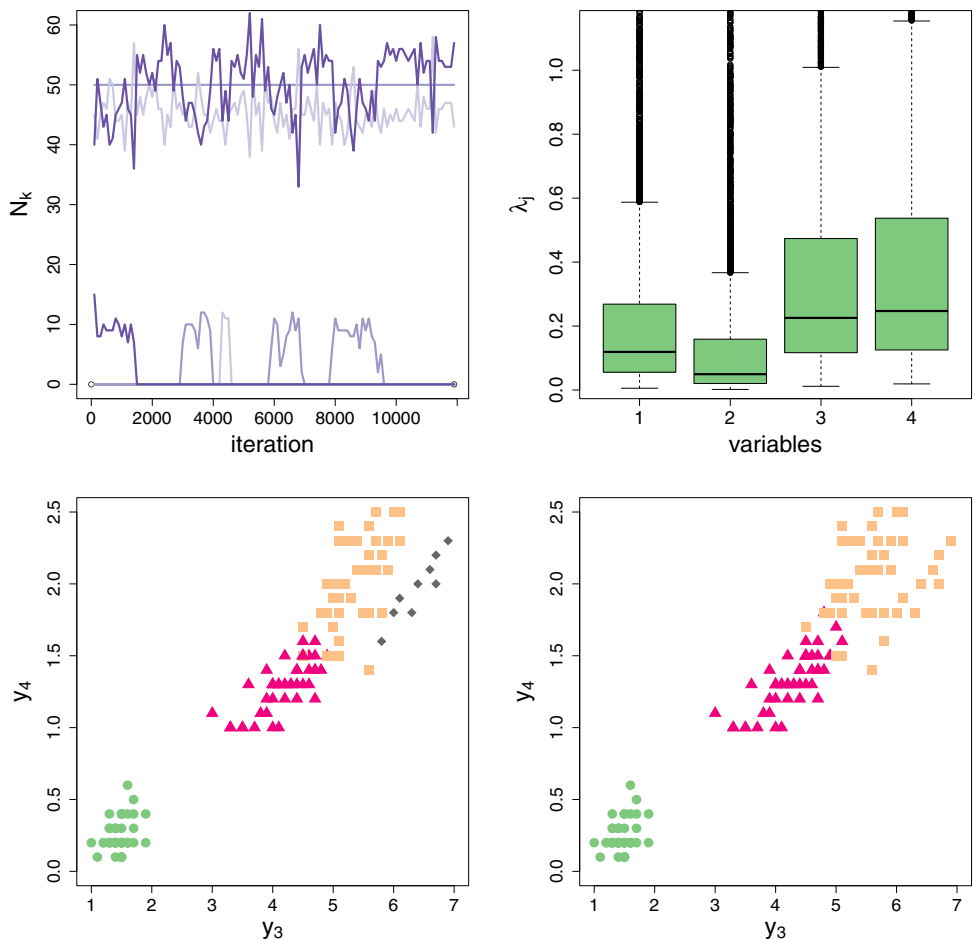

**Fig. 10** Iris data, $K = 15$: *Top row* Trace plot of number of observations allocated to the different components under the standard prior (*left-hand side*), box plots of posterior shrinkage factors $\lambda_j$, for all four variables, under the normal gamma prior (*right-hand side*). *Bottom row* estimated classification for $K_0 = 4$ under the standard prior (*left-hande side*) and true classification (*right-hande side*)

However, there is strong evidence for a fourth, non-empty component, actually not present in the true classification. Under both priors, a considerable number of draws is sam-pled from a mixture model with four non-empty components for all overfitting mixtures. We study the MCMC draws for $K = 15$ under the standard prior in more detail. On the top row, in the left-hand side plot of Fig. 10, the number of observations allocated to the different components during MCMC sampling is displayed, indicating frequent switches between 3 and 4 non-empty components. This indicates that the posterior distribution does not converge clearly to a solution with three non-empty components and that a mixture model with $K_0 = 4$ non-empty components has also considerable posterior probability. Moreover, the non-permutation rate $M_{0,\rho}$ is small for $K_0 = 4$ (0.004), indicating that the component means are well separated. If further inference is based on the draws with $K_0 = 4$ non-empty components, the obtained four-cluster solution seems to be a reasonable solution. This can be seen in Fig. 10 where scatter plots of 2 variables (petal length and petal width) under both the resulting classification for $K_0 = 4$ and the true classification are displayed. Observations of the fourth estimated component, displayed in dark grey diamonds, constitute a rather isolated group.

Regarding the identification of cluster-relevant variables, box plots of the shrinkage factors $\lambda_j$ displayed in Fig. 10 indicate that variable 2 (sepal width) is the most homogeneous variable which coincides with results in Yau and Holmes (2011).

If the number of groups is estimated by specifying an infinite mixture model, only the two data clusters are identified indicating that the infinite mixture approach implemented in package PReMiuM with the default settings aims at identifying clearly separated clusters with minimum overlap.

## 6 Discussion

In the framework of Bayesian model-based clustering, we propose a straightforward and simple strategy for simultaneous estimation of the unknown number of mixture components, component-specific parameters, classification of observations, and identification of cluster-relevant variables for multivariate Gaussian mixtures. Our approach relies on specifying sparse priors on the mixture parameters and involves only standard MCMC methods.

Estimation of the unknown number of mixture components is based on sparse finite mixtures where a sparse prior on the weights empties superfluous components during MCMC sampling and the number of true components can be estimated from the number of non-empty components. An advantage of this strategy is that model selection can be performed without computer-intensive calculations of marginal likelihoods or designing sophisticated proposals within RJMCMC. This approach works astonishingly well if the number of observations and the number of variables is not too large.

However, there are also limitations to the proposed strategy. First of all, we investigated our strategy under the assumption that the mixture components truly arise from multivariate Gaussian distributions. In order to catch non-symmetrical cluster shapes or handle outliers it would also be interesting to extend the approach to non-Gaussian component distributions, e.g. the $t$-distribution and the skew-normal distribution (see Frühwirth-Schnatter and Pyne 2010; Lee and McLachlan 2014).

We may apply sparse finite Gaussian mixtures to data from such skew or fat-tailed mixture distributions, however, in this case the posterior mixture distribution tends to fit more than one Gaussian component to represent a single non-Gaussian cluster, in particular for an increasing number of observations. As a consequence, the method is fitting Gaussian mixtures in the sense of density estimation, where the number of components is of no specific interest, and the estimated number of non-empty components no longer represents the number of distinct clusters. An important issue for further investigation is therefore how to combine mixture components, i.e. how to identify adjacent located components and merge them into one larger cluster. Several recent papers have considered the problem of merging Gaussian mixture components, see e.g. Li (2005), Baudry et al. (2010), and Hennig (2010).

To identify cluster-relevant variables, the standard prior for the component means commonly applied for multivariate Gaussian mixtures is substituted by a hierarchical shrinkage prior based on the normal gamma prior. This prior tends to fill superfluous components, since it becomes informative during MCMC sampling and superfluous components are placed in reasonable locations. We found that the normal gamma prior requires specification of a very sparse prior on the mixture weights, which is achieved by setting the hyperparameter $e_0$ of the Dirichlet prior to a very small fixed value. Under this shrinkage prior, the true number of components is recovered from overfitting finite mixtures for simulated data. Additionally, under the normal gamma prior box plots of shrinkage factors allow visual identification of cluster-relevant and homogeneous variables. Furthermore, component locations are estimated more precisely under the normal gamma than under the standard prior.

A limitation of this strategy is that explicit identification of cluster-relevant variables is based on visual detection in a post-processing step. In the presence of a huge number of variables, this strategy might be too cumbersome and an automatic approach might possibly be preferable. This could be developed similar to the automatic approaches which are for example used for explicit variable selection in a regression setting when using the Bayesian Lasso.

Finally, we note a further limitation of the proposed strategy for high-dimensional data sets. When applying sparse finite mixtures to high-dimensional data sets, Gibbs sampling with data augmentation tends to get stuck in local modes, so that superfluous components do not become empty during sampling. An issue for further studies is therefore how to improve mixing, i.e. to design well-mixing samplers, a problem also mentioned in Yau and Holmes (2011).

**Acknowledgments**   This research was supported by the Austrian Science Fund (FWF): V170-N18.

## Appendix 1: Scheme for estimating using MCMC sampling

The sampling scheme iterates the following steps:

1. Parameter simulation conditional on the classification $\mathbf{S} = (S_1, \ldots, S_N)$:

   (a) Sample

   $$\boldsymbol{\eta} \sim Dir(e_1, \ldots, e_K),$$

   where

   $$e_k = e_0 + N_k,$$

   and $N_k = \#\{i : S_i = k\}$ is the number of observations assigned to component $k$.

   (b) For $k = 1, \ldots, K$: sample

   $$\boldsymbol{\Sigma}_k^{-1} \sim \mathcal{W}(c_k, \mathbf{C}_k),$$

   where

   $$c_k = c_0 + N_k/2,$$
   $$\mathbf{C}_k = \mathbf{C}_0 + \frac{1}{2} \sum_{i : S_i = k} (\mathbf{y}_i - \boldsymbol{\mu}_k)(\mathbf{y}_i - \boldsymbol{\mu}_k)'.$$

   (c) For $k = 1, \ldots, K$: sample

   $$\boldsymbol{\mu}_k \sim \mathcal{N}(\mathbf{b}_k, \mathbf{B}_k),$$

   where

   $$\mathbf{B}_k = (\mathbf{B}_0^{-1} + N_k \boldsymbol{\Sigma}_k^{-1})^{-1},$$
   $$\mathbf{b}_k = \mathbf{B}_k(\mathbf{B}_0^{-1}\mathbf{b}_0 + \boldsymbol{\Sigma}_k^{-1} N_k \bar{\mathbf{y}}_k),$$

   and $\bar{\mathbf{y}}_k$ is the mean of the observations assigned by $\mathbf{S}$ to component $k$.

2. Classification of each observation $\mathbf{y}_i$ conditional on knowing $\boldsymbol{\mu}, \boldsymbol{\Sigma}, \boldsymbol{\eta}$:

   (a) For $i = 1, \ldots, N$: sample $S_i$ from

   $$Pr(S_i = k | \mathbf{y}_i; \boldsymbol{\mu}, \boldsymbol{\Sigma}, \boldsymbol{\eta}) \propto \eta_k f_{\mathcal{N}}(\mathbf{y}_i | \boldsymbol{\mu}_k, \boldsymbol{\Sigma}_k).$$

3. Sample hyperparameters:

   (a) Sample

   $$\mathbf{C}_0 \sim \mathcal{W}(g_0 + Kc_0, \mathbf{G}_0 + \sum_{k=1}^{K} \boldsymbol{\Sigma}_k^{-1}).$$

   (b) For a Dirichlet prior with random hyperparameter $e_0$, sample $e_0$ via a Metropolis-Hastings step from

   $$p(e_0|\boldsymbol{\eta}) \propto p(\boldsymbol{\eta}|e_0)p(e_0),$$

   see Eq. (10).

   Additionally, for the normal gamma prior:

   (c) For $j = 1, \ldots, r$, sample

   $$\lambda_j \sim \mathcal{GIG}(p_K, a_j, b_j),$$

   where

   $$p_K = \nu_1 - K/2,$$
   $$a_j = 2\nu_2,$$
   $$b_j = \sum_{k=1}^{K} (\mu_{kj} - b_{0j})^2 / R_j^2.$$

   (d) Sample

   $$\mathbf{b}_0 \sim \mathcal{N}\left(\frac{1}{K} \sum_{k=1}^{K} \boldsymbol{\mu}_k; \frac{1}{K}\mathbf{B}_0\right),$$

   where

   $$\mathbf{B}_0 = \text{Diag}(R_1^2\lambda_1, \ldots, R_r^2\lambda_r).$$

4. Random permutation of the labeling: select randomly one permutation $\rho$ of $K!$ possible permutations of $\{1, \ldots, K\}$ and substitute:

   $$\boldsymbol{\eta} = \boldsymbol{\eta}_{\rho(1,\ldots,K)},$$
   $$(\boldsymbol{\mu}_1, \ldots, \boldsymbol{\mu}_K) = (\boldsymbol{\mu}_{\rho(1)}, \ldots, \boldsymbol{\mu}_{\rho(K)}),$$
   $$(\boldsymbol{\Sigma}_1, \ldots, \boldsymbol{\Sigma}_K) = (\boldsymbol{\Sigma}_{\rho(1)}, \ldots, \boldsymbol{\Sigma}_{\rho(K)}),$$
   $$\mathbf{S} = \rho(\mathbf{S}).$$

## Appendix 2: Scheme for clustering in the point process representation

After the MCMC run, a sparse finite mixture is identified by post-processing the MCMC draws through the following steps:

1. For each iteration $m = 1, \ldots, M$ of the MCMC run, determine the number of non-empty components $K_0^{(m)}$ according to Eq. (4).

2. Estimate the true number of mixture components by $\hat{K}_0 = mode(K_0^{(m)})$, the value of the number of non-empty components occurring most often during MCMC sampling. Consider only the subsequence $M_0$ of all MCMC iterations where the number of non-empty components is exactly equal to $\hat{K}_0$.

 

3. For all $M_0$ iterations, remove the draws from empty components.

4. Arrange the remaining draws of the different component means in a matrix with $\hat{K}_0 \cdot M_0$ rows and $r$ columns. Cluster all $\hat{K}_0 \cdot M_0$ draws into $\hat{K}_0$ clusters using $K$-centroids cluster analysis (Leisch 2006) where the distance between a point and a cluster centroid is determined by the Mahalanobis distance. Details on the cluster algorithm are given in Sect. 4.2. The $K$-centroids clustering algorithm results in a classification indicating to which clusters the component-specific parameters of each single draw belong.

5. For each iteration $m$, $m = 1, \ldots, M_0$, construct a classification sequence $\rho^{(m)}$ of size $\hat{K}_0$ containing the classifications of each draw of iteration $m$.
   For $m = 1, \ldots, M_0$, check whether $\rho^{(m)}$ is a permutation of $(1, \ldots, \hat{K}_0)$. If not, remove the corresponding draws from the MCMC subsample of size $M_0$.
   The proportion of classification sequences of $M_0$ not being a permutation is denoted by $M_{0,\rho}$.

6. For the remaining $M_0(1 - M_{0,\rho})$ draws, a unique labeling is achieved by resorting the draws according to the classification sequences $\rho^{(m)}$. The resorted, identified draws can be used for further component-specific parameter inference.

To illustrate step 5, consider for instance, that for $\hat{K}_0 = 4$, for iteration $m$ a classification sequence $\rho^{(m)} = (1, 3, 4, 2)$ is obtained through the clustering procedure. That means that the draw of the first component was assigned to cluster one, the draw of the second component was assigned to cluster three and so on. In this case, the draws of this iteration are assigned to different clusters, which allows to relabel these draws. However, if a classification sequence $\rho^{(m)} = (3, 1, 2, 1)$ is obtained, then draws sampled from two different components are assigned to the same cluster and no relabeling can be performed. Thus these draws are removed from further inference because no unique labeling can be defined for this iteration.

As already observed by Frühwirth-Schnatter (2011a), all classification sequences $\rho^{(m)}$, $m = 1, \ldots, M$ obtained in step 5 are expected to be permutations, if the point process representation of the MCMC draws contains well-separated simulation clusters. If a small fraction of non-permutations $M_{0,\rho}$ is present, the posterior draws corresponding to the non-permutation sequences are removed from the $M_0$ iterations. Only the subsequence of identified draws is used for further component-specific inference. However, a high fraction $M_{0,\rho}$ indicates that in the point process representation clusters are highly overlapping. This typically happens if the selected mixture model with $\hat{K}_0$ components is overfitting, see Frühwirth-Schnatter (2011a).

## References

Anderson, E.: The Irises of the Gaspé Peninsula. Bull. Am. Iris Soc. **59**, 2–5 (1935)

Armagan, A., Dunson, D., Clyde, M.: Generalized beta mixtures of Gaussians. In: Shawe-Taylor, J., Zemel, R., Bartlett, P., Pereira, F., Weinberger, K. (eds.) Advances in Neural Information Processing Systems (NIPS) **24**, pp. 523–531, Curran Associates, Inc., (2011)

Banfield, J.D., Raftery, A.E.: Model-based Gaussian and non-Gaussian clustering. Biometrics **49**, 803–821 (1993)

Baudry, J., Raftery, A.E., Celeux, G., Lo, K., Gottardo, R.: Combining mixture components for clustering. J. Comput. Gr. Stat. **19**, 332–353 (2010)

Bensmail, H., Celeux, G., Raftery, A.E., Robert, C.P.: Inference in model-based cluster analysis. Stat. Comput. **7**, 1–10 (1997)

Biernacki, C., Celeux, G., Govaert, G.: Assessing a mixture model for clustering with the integrated completed likelihood. IEEE Trans. Pattern Anal. Mach. Intell. **22**(7), 719–725 (2000)

Campbell, N., Mahon, R.: A multivariate study of variation in two species of rock crab of genus Leptograpsus. Austr. J. Zool. **22**, 417–425 (1974)

Celeux, G.: Bayesian inference for mixture: the label switching problem. In: Green, P.J., Rayne, R. (eds.) COMPSTAT 98, pp. 227–232. Physica, Heidelberg (1998)

Celeux, G., Hurn, M., Robert, C.P.: Computational and inferential difficulties with mixture posterior distributions. J. Am. Stat. Assoc. **95**, 957–970 (2000)

Celeux, G., Forbes, F., Robert, C.P., Titterington, D.M.: Deviance information criteria for missing data models. Bayesian Anal. **1**(4), 651–674 (2006)

Chung, Y., Dunson, D.: Nonparametric Bayes conditional distribution modeling with variable selection. J. Am. Stat. Assoc. **104**, 1646–1660 (2009)

Dasgupta, A., Raftery, A.E.: Detecting features in spatial point processes with clutter via model-based clustering. J. Am. Stat. Assoc. **93**(441), 294–302 (1998)

Dean, N., Raftery, A.E.: Latent class analysis variable selection. Ann. Inst. Stat. Math. **62**, 11–35 (2010)

Dellaportas, P., Papageorgiou, I.: Multivariate mixtures of normals with unknown number of components. Stat. Comput. **16**, 57–68 (2006)

Diebolt, J., Robert, C.P.: Estimation of finite mixture distributions through Bayesian sampling. J. R. Stat. Soc. B **56**, 363–375 (1994)

Fisher, R.: The use of multiple measurements in taxonomic problems. Ann. Eugenics **7**(2), 179–188 (1936)

Frühwirth-Schnatter, S.: Markov chain Monte Carlo estimation of classical and dynamic switching and mixture models. J. Am. Stat. Assoc. **96**(453), 194–209 (2001)

Frühwirth-Schnatter, S.: Estimating marginal likelihoods for mixture and Markov switching models using bridge sampling techniques. Econ. J. **7**, 143–167 (2004)

Frühwirth-Schnatter, S.: Finite Mixture and Markov Switching Models. Springer-Verlag, New York (2006)

Frühwirth-Schnatter, S.: Label switching under model uncertainty. In: Mengerson, K., Robert, C., Titterington, D. (eds.) Mixtures: Estimation and Application, pp. 213–239. Wiley, New York (2011a)

Frühwirth-Schnatter, S.: Panel data analysis - a survey on model-based clustering of time series. Adv. Data Anal. Classif. **5**(4), 251–280 (2011b)

Frühwirth-Schnatter, S., Kaufmann, S.: Model-based clustering of multiple time series. J. Bus. Econ. Stat. **26**(1), 78–89 (2008)

Frühwirth-Schnatter, S., Pyne, S.: Bayesian inference for finite mixtures of univariate and multivariate skew-normal and skew-t distributions. Biostatistics **11**(2), 317–336 (2010)

Geweke, J.: Interpretation and inference in mixture models: simple MCMC works. Comput. Stat. Data Anal. **51**, 3529–3550 (2007)

Griffin, J.E., Brown, P.J.: Inference with normal-gamma prior distributions in regression problems. Bayesian Anal. **5**(1), 171–188 (2010)

Grün, B., Leisch, F.: Dealing with label switching in mixture models under genuine multimodality. J. Multivar. Anal. **100**(5), 851–861 (2009)

Handcock, M.S., Raftery, A.E., Tantrum, J.M.: Model-based clustering for social networks. J. R. Stat. Soc. A **170**(2), 301–354 (2007)

Hennig, C.: Methods for merging Gaussian mixture components. Adv. Data Anal. Classif. **4**, 3–34 (2010)

Ishwaran, H., James, L.F., Sun, J.: Bayesian model selection in finite mixtures by marginal density decompositions. J. Am. Stat. Assoc. **96**(456), 1316–1332 (2001)

Jasra, A., Holmes, C.C., Stephens, D.A.: Markov chain Monte Carlo methods and the label switching problem in Bayesian mixture modeling. Stat. Sci. **20**(1), 50–67 (2005)

Juárez, M.A., Steel, M.F.J.: Model-based clustering of non-Gaussian panel data based on skew-t distributions. J. Bus. Econ. Stat. **28**(1), 52–66 (2010)

Kaufman, L., Rousseeuw, P.J.: Finding Groups in Data: An Introduction to Cluster Analysis. Wiley, New York (1990)

Kim, S., Tadesse, M.G., Vannucci, M.: Variable selection in clustering via Dirichlet process mixture models. Biometrika **93**(4), 877–893 (2006)

Kundu, S., Dunson, D.B.: Bayes variable selection in semiparametric linear models. J. Am. Stat. Assoc. **109**(505), 437–447 (2014)

Lee, H., Li, J.: Variable selection for clustering by separability based on ridgelines. J. Comput. Gr. Stat. **21**(2), 315–337 (2012)

Lee, S., McLachlan, G.J.: Finite mixtures of multivariate skew t-distributions: some recent and new results. Stat. Comput. **24**(2), 181–202 (2014)

Leisch, F.: A toolbox for $K$-centroids cluster analysis. Comput. Stat. Data Anal. **51**(2), 526–544 (2006)

Li, J.: Clustering based on a multi-layer mixture model. J. Comput. Gr. Stat. **14**, 547–568 (2005)

Lian, H.: Sparse Bayesian hierarchical modeling of high-dimensional clustering problems. J. Multivar. Anal. **101**(7), 1728–1737 (2010)

Liverani, S., Hastie, D.I., Papathomas, M., Richardson, S.: PReMiuM: An R package for profile regression mixture models using Dirichlet processes, arXiv preprint arXiv:1303.2836 (2013)

Maugis, C., Celeux, G., Martin-Magniette, M.L.: Variable selection for clustering with Gaussian mixture models. Biometrics **65**(3), 701–709 (2009)

McLachlan, G.J., Peel, D.: Finite Mixture Models. Wiley series in probability and statistics. Wiley, New York (2000)

McLachlan, G.J., Bean, R.W., Peel, D.: A mixture-model based approach to the clustering of microarray expression data. Bioinformatics **18**, 413–422 (2002)

McNicholas, P.D., Murphy, T.B.: Parsimonious Gaussian mixture models. Stat. Comput **18**(3), 285–296 (2008)

McNicholas, P.D., Murphy, T.B.: Model-based clustering of longitudinal data. Can. J. Stat. **38**(1), 153–168 (2010)

Molitor, J., Papathomas, M., Jerrett, M., Richardson, S.: Bayesian profile regression with an application to the national survey of children's health. Biostatistics **11**(3), 484–498 (2010)

Nobile, A.: On the posterior distribution of the number of components in a finite mixture. Ann. Stat. **32**, 2044–2073 (2004)

Pan, W., Shen, X.: Penalized model-based clustering with application to variable selection. J. Mach. Learn. Res. **8**, 1145–1164 (2007)

Park, T., Casella, G.: The Bayesian Lasso. J. Am. Stat. Assoc. **103**(482), 681–686 (2008)

Polson, N.G., Scott, J.G.: Shrink globally, act locally: sparse Bayesian regularization and prediction. In: Bernardo, J., Bayarri, M., Berger, J., Dawid, A., Heckerman, D., Smith, A., West, M. (eds.) Bayesian Statistics, vol. 9, pp. 501–523. Oxford University Press, Oxford (2010)

Raftery, A.E., Dean, N.: Variable selection for model-based clustering. J. Am. Stat. Assoc. **101**(473), 168–178 (2006)

Richardson, S., Green, P.J.: On Bayesian analysis of mixtures with an unknown number of components. J. R. Stat. Soc. B **59**(4), 731–792 (1997)

Rousseau, J., Mengersen, K.: Asymptotic behaviour of the posterior distribution in overfitted mixture models. J. R. Stat. Soc. B **73**(5), 689–710 (2011)

Sperrin, M., Jaki, T., Wit, E.: Probabilistic relabelling strategies for the label switching problem in Bayesian mixture models. Stat. Comput. **20**(3), 357–366 (2010)

Stephens, M.: Bayesian methods for mixtures of normal distributions. Ph.D. thesis, University of Oxford (1997)

Stephens, M.: Dealing with label switching in mixture models. J. R. Stat. Soc. B **62**, 795–809 (2000)

Stingo, F.C., Vannucci, M., Downey, G.: Bayesian wavelet-based curve classification via discriminant analysis with Markov random tree priors. Statistica Sinica **22**(2), 465 (2012)

Tadesse, M.G., Sha, N., Vanucci, M.: Bayesian variable selection in clustering high-dimensional data. J. Am. Stat. Assoc. **100**(470), 602–617 (2005)

Venables, W.N., Ripley, B.D.: Modern Applied Statistics with S, 4th edn. Springer-Verlag, New York (2002)

Wang, S., Zhu, J.: Variable selection for model-based high-dimensional clustering and its application to microarray data. Biometrics **64**(2), 440–448 (2008)

Xie, B., Pan, W., Shen, X.: Variable selection in penalized model-based clustering via regularization on grouped parameters. Biometrics **64**(3), 921–930 (2008)

Yao, W., Lindsay, B.G.: Bayesian mixture labeling by highest posterior density. J. Am. Stat. Assoc. **104**, 758–767 (2009)

Yau, C., Holmes, C.: Hierarchical Bayesian nonparametric mixture models for clustering with variable relevance determination. Bayesian Anal. **6**(2), 329–352 (2011)

Yeung, K.Y., Fraley, C., Murua, A., Raftery, A.E., Ruzzo, W.L.: Model-based clustering and data transformations for gene expression data. Bioinformatics **17**, 977–987 (2001)